# Thermodynamics of Radiation Pressure and Photon Momentum

Masud Mansuripur[†] and Pin Han[‡]

[†]College of Optical Sciences, the University of Arizona, Tucson, Arizona, USA

[‡]Graduate Institute of Precision Engineering, National Chung Hsing University, Taichung, Taiwan

**Abstract**. Theoretical analyses of radiation pressure and photon momentum in the past 150 years have focused almost exclusively on classical and/or quantum theories of electrodynamics. In these analyses, Maxwell's equations, the properties of polarizable and/or magnetizable material media, and the stress tensors of Maxwell, Abraham, Minkowski, Chu, and Einstein-Laub have typically played prominent roles [1-9]. Each stress tensor has subsequently been manipulated to yield its own expressions for the electromagnetic (EM) force, torque, energy, and linear as well as angular momentum densities of the EM field. This paper presents an alternative view of radiation pressure from the perspective of thermal physics, invoking the properties of blackbody radiation in conjunction with empty as well as gas-filled cavities that contain EM energy in thermal equilibrium with the container's walls. In this type of analysis, Planck's quantum hypothesis, the spectral distribution of the trapped radiation, the entropy of the photon gas, and Einstein's *A* and *B* coefficients play central roles.

**1. Introduction**. At a fundamental level, many properties of the electromagnetic (EM) radiation are intimately tied to the basic tenets of thermodynamics, which is essentially the science that seeks answers to the question: "*What must the laws of nature be like so that it is impossible to construct a perpetual motion machine of either the first or the second kind?*" This question refers to the empirical fact that perpetual motion machines have never been observed that could violate the first principle (i.e., energy cannot be added or destroyed in an isolated system), or contradict the second principle (i.e., entropy always increases for spontaneous processes in an isolated system) [10]. According to Albert Einstein (1879-1955), "*the science of thermodynamics seeks by analytical means to deduce necessary conditions, which separate events have to satisfy, from the universally experienced fact that perpetual motion is impossible.*" In his monumental work on EM radiation, Einstein, who was well-versed in thermal physics, drew profound original conclusions from the principles of statistical mechanics and thermodynamics. Einstein's enthusiasm and admiration for this branch of physical science is palpable in the following reflection on his early career: "*A law is more impressive the greater the simplicity of its premises, the more different are the kinds of things it relates, and the more extended its range of applicability. Therefore the deep impression that classical thermodynamics made upon me. It is the only physical theory of universal content, which I am convinced, that within the framework of applicability of its basic concepts will never be overthrown.*"

The goal of the present paper is to review the physics of blackbody radiation and to draw a few conclusions concerning radiation pressure from the principles of thermodynamics and statistical physics. Section 2 provides a detailed derivation of Planck's blackbody radiation formula from a combination of classical and quantum electrodynamics. In Sec. 3 we relate Planck's formula to the Stefan-Boltzmann law of blackbody radiation. Section 4 describes the radiation pressure inside an evacuated chamber whose walls are maintained at a constant temperature $T$. In Sec. 5 we show that the Stefan-Boltzmann law is unaffected when the blackbody chamber is filled with a gas of refractive index $n(\omega)$. Section 6 explains why the EM momentum density inside a gas-filled chamber at temperature $T$ is given by the Abraham formula, despite the fact that the radiation pressure on the walls of the chamber appears to conform with the Minkowski formulation. Fluctuations of thermal radiation are the subject of Sec. 7, while different approaches to deriving an expression for the entropy of the 'photon gas' form the subject of Sec. 8. In Sec. 9 we return to Planck's formula and derive it from a different perspective, this time from a combination of classical electrodynamics and the quantum theory of material media acting as a collection of harmonic oscillators. Section 10 addresses the intimate connection between Einstein's *A* and *B* coefficients and Planck's blackbody radiation formula. Certain implications of the second law of thermodynamics for radiation pressure on various surfaces are the subject of Sec.11. In Sec.12 we reconstruct an argument from Einstein's early papers to show that quantization of radiation is necessary if a hypothetical reflective foil floating inside a radiation-filled (but otherwise empty) chamber at temperature $T$ is to behave in accordance with the equipartition theorem of statistical physics. The final section is devoted to a summary of the results and a few concluding remarks.

**2. Blackbody radiation**. With reference to Fig.1(a), consider a hollow rectangular parallelepiped box of dimensions $L_x \times L_y \times L_z$, with perfectly reflecting interior walls, filled with EM radiation of all admissible frequencies. The boundary conditions allow only certain $k$-vectors (and, considering that $|\boldsymbol{k}| = \omega/c$, only certain frequencies) in the form of

$$\boldsymbol{k} = (\pi n_x/L_x)\hat{\boldsymbol{x}} + (\pi n_y/L_y)\hat{\boldsymbol{y}} + (\pi n_z/L_z)\hat{\boldsymbol{z}}. \tag{1}$$

In the above equation, $(n_x, n_y, n_z)$ are arbitrary integers, which may be positive or negative but not zero. For every choice of these three integers, we construct a standing wave within the confines of the hollow cavity by superposing the eight plane-waves associated with the combinations $(\pm n_x, \pm n_y, \pm n_z)$.

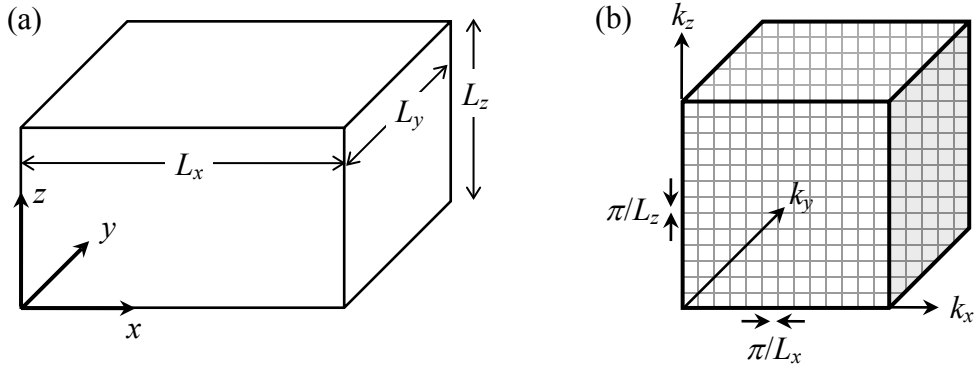

**Fig.1**. (a) Rectangular parallelepiped chamber having dimensions $L_x \times L_y \times L_z$. The internal walls of the cavity are perfectly reflecting, thus allowing standing EM waves of certain frequencies to survive indefinitely within the cavity. (b) Allowed $k$-vectors of the resonant modes of the chamber form a discrete mesh in the $k$-space, with each volume element, having dimensions $(\pi/L_x) \times (\pi/L_y) \times (\pi/L_z)$, containing exactly one $k$-vector.

Note that the constraint $\boldsymbol{k} \cdot \boldsymbol{E}_0 = 0$ imposed by Maxwell's 1st equation must be reflected in the choice of signs for the various $E$-field components. Denoting the $E$-field components by $E_{x0} = |E_{x0}|\exp(\mathrm{i}\phi_{x0})$, $E_{y0} = |E_{y0}|\exp(\mathrm{i}\phi_{y0})$, and $E_{z0} = |E_{z0}|\exp(\mathrm{i}\phi_{z0})$, we write

$$\boldsymbol{E}(\boldsymbol{r},t) = \mathrm{Real} \sum_{m=1}^{8} \boldsymbol{E}_0^{(m)} \exp\{\mathrm{i}[\boldsymbol{k}^{(m)} \cdot \boldsymbol{r} - \omega t]\}$$

$$= \mathrm{Re}\{(E_{x0}\hat{\boldsymbol{x}} + E_{y0}\hat{\boldsymbol{y}} + E_{z0}\hat{\boldsymbol{z}})\exp[\mathrm{i}(k_x x + k_y y + k_z z - \omega t)] - (E_{x0}\hat{\boldsymbol{x}} + E_{y0}\hat{\boldsymbol{y}} - E_{z0}\hat{\boldsymbol{z}})\exp[\mathrm{i}(k_x x + k_y y - k_z z - \omega t)]$$

$$-(E_{x0}\hat{\boldsymbol{x}} - E_{y0}\hat{\boldsymbol{y}} + E_{z0}\hat{\boldsymbol{z}})\exp[\mathrm{i}(k_x x - k_y y + k_z z - \omega t)] + (E_{x0}\hat{\boldsymbol{x}} - E_{y0}\hat{\boldsymbol{y}} - E_{z0}\hat{\boldsymbol{z}})\exp[\mathrm{i}(k_x x - k_y y - k_z z - \omega t)]$$

$$+(E_{x0}\hat{\boldsymbol{x}} - E_{y0}\hat{\boldsymbol{y}} - E_{z0}\hat{\boldsymbol{z}})\exp[\mathrm{i}(-k_x x + k_y y + k_z z - \omega t)] - (E_{x0}\hat{\boldsymbol{x}} - E_{y0}\hat{\boldsymbol{y}} + E_{z0}\hat{\boldsymbol{z}})\exp[\mathrm{i}(-k_x x + k_y y - k_z z - \omega t)]$$

$$-(E_{x0}\hat{\boldsymbol{x}} + E_{y0}\hat{\boldsymbol{y}} - E_{z0}\hat{\boldsymbol{z}})\exp[\mathrm{i}(-k_x x - k_y y + k_z z - \omega t)] + (E_{x0}\hat{\boldsymbol{x}} + E_{y0}\hat{\boldsymbol{y}} + E_{z0}\hat{\boldsymbol{z}})\exp[\mathrm{i}(-k_x x - k_y y - k_z z - \omega t)]\}$$

$$= -8|E_{x0}|\cos(k_x x)\sin(k_y y)\sin(k_z z)\cos(\omega t - \phi_{x0})\hat{\boldsymbol{x}} - 8|E_{y0}|\sin(k_x x)\cos(k_y y)\sin(k_z z)\cos(\omega t - \phi_{y0})\hat{\boldsymbol{y}}$$

$$-8|E_{z0}|\sin(k_x x)\sin(k_y y)\cos(k_z z)\cos(\omega t - \phi_{z0})\hat{\boldsymbol{z}}. \tag{2}$$

It is clear that the $E$-field components parallel to the six walls of the chamber vanish at each wall. Where the $E$-field is perpendicular to a surface, it does not vanish at the surface, because the relevant boundary condition is satisfied by the presence of electric charges at these surfaces.

Next we calculate the distribution of the $H$-field inside the chamber. The relationship between the components of $\boldsymbol{E}_0$, $\boldsymbol{H}_0$ and $\boldsymbol{k}$ is given by Maxwell's 3$^{rd}$ equation as $\boldsymbol{k} \times \boldsymbol{E}_0 = \mu_0 \omega \boldsymbol{H}_0$ [2,11]. Thus, changes in the signs of the various components of $\boldsymbol{E}_0$ and $\boldsymbol{k}$ must be reflected in the signs of the components of $\boldsymbol{H}_0$ for each plane-wave. Maxwell's 4$^{th}$ equation requires that $\boldsymbol{k} \cdot \boldsymbol{H}_0 = 0$, but this is automatically satisfied when $\boldsymbol{H}_0$ is derived from Maxwell's 3$^{rd}$ equation. Denoting the $H$-field components by $H_{x0} = |H_{x0}| \exp(i\varphi_{x0})$, etc., we will have

$$\boldsymbol{H}(\boldsymbol{r},t) = \text{Real} \sum_{m=1}^{8} \boldsymbol{H}_0^{(m)} \exp\{i[\boldsymbol{k}^{(m)} \cdot \boldsymbol{r} - \omega t]\}$$

$$= \text{Re}\{(H_{x0}\hat{\boldsymbol{x}} + H_{y0}\hat{\boldsymbol{y}} + H_{z0}\hat{\boldsymbol{z}}) \exp[i(k_x x + k_y y + k_z z - \omega t)] + (H_{x0}\hat{\boldsymbol{x}} + H_{y0}\hat{\boldsymbol{y}} - H_{z0}\hat{\boldsymbol{z}}) \exp[i(k_x x + k_y y - k_z z - \omega t)]$$

$$+ (H_{x0}\hat{\boldsymbol{x}} - H_{y0}\hat{\boldsymbol{y}} + H_{z0}\hat{\boldsymbol{z}}) \exp[i(k_x x - k_y y + k_z z - \omega t)] + (H_{x0}\hat{\boldsymbol{x}} - H_{y0}\hat{\boldsymbol{y}} - H_{z0}\hat{\boldsymbol{z}}) \exp[i(k_x x - k_y y - k_z z - \omega t)]$$

$$- (H_{x0}\hat{\boldsymbol{x}} - H_{y0}\hat{\boldsymbol{y}} - H_{z0}\hat{\boldsymbol{z}}) \exp[i(-k_x x + k_y y + k_z z - \omega t)] - (H_{x0}\hat{\boldsymbol{x}} - H_{y0}\hat{\boldsymbol{y}} + H_{z0}\hat{\boldsymbol{z}}) \exp[i(-k_x x + k_y y - k_z z - \omega t)]$$

$$- (H_{x0}\hat{\boldsymbol{x}} + H_{y0}\hat{\boldsymbol{y}} - H_{z0}\hat{\boldsymbol{z}}) \exp[i(-k_x x - k_y y + k_z z - \omega t)] - (H_{x0}\hat{\boldsymbol{x}} + H_{y0}\hat{\boldsymbol{y}} + H_{z0}\hat{\boldsymbol{z}}) \exp[i(-k_x x - k_y y - k_z z - \omega t)]$$

$$= 8|H_{x0}| \sin(k_x x) \cos(k_y y) \cos(k_z z) \sin(\omega t - \varphi_{x0}) \hat{\boldsymbol{x}} + 8|H_{y0}| \cos(k_x x) \sin(k_y y) \cos(k_z z) \sin(\omega t - \varphi_{y0}) \hat{\boldsymbol{y}}$$

$$+ 8|H_{z0}| \cos(k_x x) \cos(k_y y) \sin(k_z z) \sin(\omega t - \varphi_{z0}) \hat{\boldsymbol{z}}. \tag{3}$$

The $H$-field components that are perpendicular to a given surface are seen to vanish at that surface, thus satisfying the boundary condition for the perpendicular $B$-field. The tangential $H$-fields, however, do not have to vanish at various surfaces, as they are matched by the induced surface currents [2,11].

Each of the above standing waves constitutes a mode of the resonator (or cavity). In fact, for each triplet of positive integers $(n_x, n_y, n_z)$, there exist *two* resonator modes, because each standing wave can have two independent states of polarization. This is due to the fact that the $E$-field amplitudes $(E_{x0}, E_{y0}, E_{z0})$ are constrained by Maxwell's 1$^{st}$ equation to satisfy $\boldsymbol{k} \cdot \boldsymbol{E}_0 = 0$ and, therefore, the initial choice of $\boldsymbol{E}_0$ represents two (rather than three) degrees of freedom for a plane-wave whose $k$-vector is already fixed by the choice of $(n_x, n_y, n_z)$.

The volume occupied by each cavity mode in the $k$-space is $\delta^3 = \pi^3/(L_x L_y L_z)$; see Eq.(1). Using the dispersion relation $k_x^2 + k_y^2 + k_z^2 = (\omega/c)^2$ for plane-waves in vacuum, we compute the number of modes in the $k$-space confined to a thin spherical shell of radius $k = \omega/c$ and thickness $dk = d\omega/c$ to be $4\pi k^2 dk/\delta^3 = 4L_x L_y L_z \omega^2 d\omega/(\pi^2 c^3)$. Only one octant of the spherical shell, however, corresponds to positive values of $(n_x, n_y, n_z)$, which represent the various standing waves. Therefore, the above number of modes must be divided by 8, then multiplied by 2 to account for the two polarization states of each standing wave. The number of modes thus obtained may be further normalized by the volume $L_x L_y L_z$ of the cavity, as we are primarily interested in the energy-density (i.e., energy per unit volume) of the resonant EM waves. The number-density of all allowed modes in the frequency interval $(\omega, \omega + d\omega)$ is thus given by $\omega^2 d\omega/(\pi^2 c^3)$.

Each resonant cavity mode associated with frequency $\omega$ may be occupied by $m$ identical photons of energy $\hbar\omega$ with a probability proportional to $\exp(-m\hbar\omega/k_B T)$. Here $m$ is an arbitrary integer $(0, 1, 2, \cdots)$, $\hbar = 1.054572 \times 10^{-34}$ *Joule·sec* is Planck's reduced constant, $k_B = 1.38065 \times 10^{-23}$ *Joule/°K* is Boltzmann's constant, and $T$ is the absolute temperature of the cavity. (The cavity walls, being perfect reflectors, cannot emit any radiation into the chamber. One must assume that a material body of temperature $T$ has resided in the cavity for a sufficiently long time to have established thermal equilibrium with the EM field, before being removed without disturbing the radiation state.) The occupation probability of a given mode of frequency $\omega$ at temperature $T$ is $\exp(-m\hbar\omega/k_B T)$ normalized by the so-called partition function $Z(\omega, T)$, which is defined as

$$Z(\omega, T) = \sum_{m=0}^{\infty} \exp(-m\hbar\omega/k_B T) = \frac{1}{1 - \exp(-\hbar\omega/k_B T)}. \tag{4}$$

The derivative with respect to $\omega$ of the natural logarithm of the partition function yields the average photon number occupancy for the mode under consideration, as follows:

$$\frac{\partial[\ln Z(\omega,T)]}{\partial \omega} = \frac{\partial Z/\partial \omega}{Z} = -(\hbar/k_B T)\frac{\sum_{m=0}^{\infty} m \exp(-m\hbar\omega/k_B T)}{\sum_{m=0}^{\infty} \exp(-m\hbar\omega/k_B T)} = -\hbar\langle m\rangle/(k_B T). \qquad (5)$$

The desired derivative is readily found from Eq.(4) to be

$$\frac{\partial[\ln Z(\omega,T)]}{\partial \omega} = -\frac{\partial[1-\exp(-\hbar\omega/k_B T)]/\partial\omega}{1-\exp(-\hbar\omega/k_B T)} = -\frac{\hbar/(k_B T)}{\exp(\hbar\omega/k_B T)-1}. \qquad (6)$$

Comparing Eq.(5) with Eq.(6), we find

$$\langle m\rangle = \frac{1}{\exp(\hbar\omega/k_B T)-1}. \qquad (7)$$

As expected, the occupancy of modes whose $\hbar\omega$ is below or comparable to $k_B T$ is large, whereas modes whose photon energy $\hbar\omega$ far exceeds $k_B T$ are sparsely populated. The energy density of blackbody radiation in the frequency interval $(\omega, \omega + d\omega)$ may now be obtained by multiplying the mode number density $\omega^2 d\omega/(\pi^2 c^3)$ with the average photon energy per mode, $\langle m\rangle\hbar\omega$. The energy density (per unit volume per unit angular frequency) is thus found to be [11-13]

$$\mathcal{E}(\omega,T) = \frac{\hbar\omega^3}{\pi^2 c^3[\exp(\hbar\omega/k_B T)-1]}. \qquad (8)$$

This is the well-known blackbody radiation energy-density (energy per unit volume per unit angular frequency) that has been associated with the name of Max Planck. Integrating $\mathcal{E}(\omega,T)$ over all frequencies $\omega$ ranging from 0 to $\infty$, yields a total energy-density (per unit volume) proportional to $T^4$, in agreement with the Stefan-Boltzmann law.

$$u(T) = \int_0^{\infty}\mathcal{E}(\omega,T)d\omega = \int_0^{\infty}\frac{\hbar\omega^3}{\pi^2 c^3[\exp(\hbar\omega/k_B T)-1]}d\omega = \left(\frac{k_B^4 T^4}{\pi^2 c^3 \hbar^3}\right)\int_0^{\infty}\frac{x^3 dx}{\exp(x)-1} \overset{\text{Gradshteyn \& Ryzhik 3.411-17 [26]}}{=} \left(\frac{\pi^2 k_B^4}{15 c^3 \hbar^3}\right)T^4. \qquad (9)$$

At low frequencies or high temperatures, the exponential factor in the denominator on the right-hand-side of Eq.(8) may be approximated as $1 + (\hbar\omega/k_B T)$, thus reducing the energy-density formula to $\mathcal{E}(\omega,T) \cong \omega^2 k_B T/(\pi^2 c^3)$. This is the Rayleigh-Jeans formula for the energy density of blackbody radiation, derived under the assumption that each cavity mode, in accordance with the equi-partition theorem [11-14], partakes equally of the available energy in the amount of $\tfrac{1}{2}k_B T$ per available degree of freedom. Since each cavity mode was deemed to behave as a harmonic oscillator holding $\tfrac{1}{2}k_B T$ of kinetic energy and $\tfrac{1}{2}k_B T$ of potential energy (i.e., electric and magnetic field energies of the mode), a total energy of $k_B T$ was assigned to each mode.

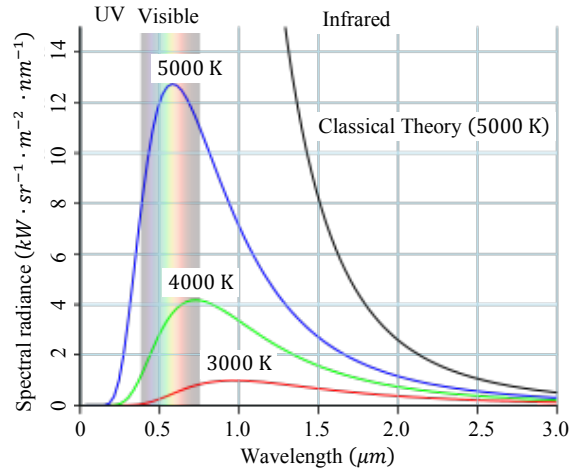

**Fig. 2**. Plots of the blackbody energy density versus the wavelength for three different temperatures. Also shown is a plot of the Rayleigh-Jeans energy density at $T = 5000$K.

The Rayleigh-Jeans theory gave the correct explanation for the low-frequency end of the blackbody spectrum, but it hugely over-estimated the high-energy (i.e., short-wavelength) part of the spectrum. This major disagreement between the prevailing theory of the day and experimental facts came to be known as the *ultraviolet catastrophe*, which was resolved only after Max Planck advanced his quantum hypothesis in the closing days of the 19th century.

**3. Energy emanating from the surface of a blackbody.** With reference to Fig.3, consider a small aperture of area $\delta A$ on a wall of the blackbody examined in the preceding section. During a short time interval $\delta t$, a ray (i.e., $k$-vector) emerging at an angle $\theta$ relative to the $x$-axis brings out $\tfrac{1}{2}c\delta t\delta A \cos\theta$ times the energy-density associated with that $k$-vector. The reason for the ½ factor is that for every $k$-vector with a positive $k_x$ that leaves the box, there exists one with a negative $k_x$, which would *not* come out of the aperture. The remaining factor, $c\delta t\delta A \cos\theta$, is the volume of a parallelepiped with base $\delta A$ and height $c\delta t$ along the emergent ray. Now, the average value of $\cos\theta$ among all the $k$-vectors in the first octant of the $k_x k_y k_z$-space is $\int_0^{\pi/2} \cos\theta \sin\theta \, d\theta = \tfrac{1}{2}$. The energy emanating from the aperture per unit area per unit time per unit angular frequency is thus given by

$$I(\omega, T) = \tfrac{1}{4} c \mathcal{E}(\omega, T) = \frac{\hbar \omega^3}{4\pi^2 c^2 [\exp(\hbar\omega/k_B T) - 1]}. \tag{10}$$

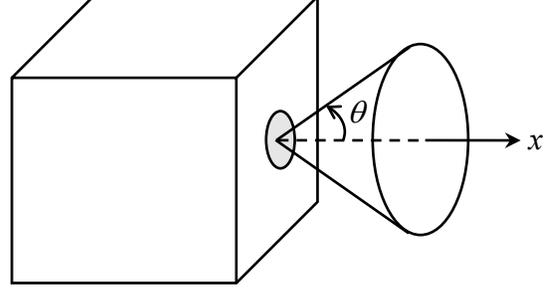

Fig.3. Radiation emerging from a small aperture of area $\delta A$ on the surface of a blackbody box.

When integrated over all frequencies, the above formula yields

$$\int_0^\infty I(\omega, T) d\omega = \left(\frac{\pi^2 k_B^4}{60 c^2 \hbar^3}\right) T^4 = \sigma T^4. \tag{11}$$

The numerical value of $\sigma$, the Stefan-Boltzmann constant, is $5.670373(21) \times 10^{-8}$ J/(s·m²·K⁴). A direct derivation of the Stefan-Boltzmann law from thermodynamics principles is given in Appendix A.

**4. Radiation pressure inside the blackbody chamber.** The momentum-density of a plane-wave propagating in free space along a given $k$-vector is equal to the energy-density divided by the speed of light, $c$. The direction of the momentum, of course, is aligned with the $k$-vector. Consider a small volume in the shape of a parallelepiped adjacent to a wall inside the blackbody of Fig.1. The volume of the parallelepiped is $c\delta t \delta A \cos\theta$, where $\delta A$ is the base area, $\delta t$ is a short time interval, and $\theta$ is the angle between the surface-normal and the particular $k$-vector under consideration. For each $k$-vector pointing toward the wall, there is an identical $k$-vector pointing away, so only half the momentum content of the parallelepiped bounces off its adjacent wall. The change of momentum upon reflection from the wall involves multiplication by the factor $2\cos\theta$, because the perpendicular component of the incident momentum changes sign upon reflection, whereas the tangential component remains unchanged. Thus, the net momentum transferred to the wall during the time interval $\delta t$ is $\tfrac{1}{2}(2c\delta t \delta A \cos^2\theta)$ times the momentum density. Dividing this transferred momentum by $\delta t$ yields the force, and further division by $\delta A$ yields the pressure on the wall, so the pressure exerted by each mode (corresponding to a given $k$-vector) is the energy density of the mode multiplied by $\cos^2\theta$. Since the average $\cos^2\theta$ over all the $k$-vectors in the first octant of the $k_x k_y k_z$-space is $\int_0^{\pi/2} \cos^2\theta \sin\theta \, d\theta = \tfrac{1}{3}$, the radiation pressure inside the chamber will be

$$p(T) = \tfrac{1}{3} u(T). \tag{12}$$

An alternative determination of the pressure inside the chamber involves calculating the EM force exerted on each mirror surface. With reference to Fig.1, consider the $L_y \times L_z$ surface located at $x = L_x$. The perpendicular component of the $E$-field acting on this surface is given in Eq.(2), namely,

$$E_x(x = L_x, y, z, t) = -8|E_{x0}| \cos(\pi n_x) \sin(k_y y) \sin(k_z z) \cos(\omega t - \phi_{x0}). \tag{13}$$

The induced surface-charge-density on the mirror at $x = L_x$ will thus be $\sigma_s = -\varepsilon_0 E_x$, yielding a force per unit area $F_x = \tfrac{1}{2}\sigma_s E_x = -\tfrac{1}{2}\varepsilon_0 E_x^2 = -32\varepsilon_0 |E_{x0}|^2 \sin^2(k_y y) \sin^2(k_z z) \cos^2(\omega t - \phi_{x0})$. The

factor ½ in the above equation arises from the fact that the average $E$-field acting on the surface charge is only half the $E$-field at the mirror surface, as the $E$-field inside the perfectly conducting wall is zero. Time averaging as well as averaging over the $y$ and $z$ coordinates then yields $\langle F_x \rangle = -4\varepsilon_0 |E_{x0}|^2$. In similar fashion, we find the force of the magnetic field on the mirror located at $x = L_x$ by calculating first the $H$-field from Eq.(3) followed by the induced surface current-density $\boldsymbol{J}_s$, as follows:

$$\boldsymbol{H}(x = L_x, y, z, t) = 8\cos(\pi n_x) \big[ |H_{y_0}| \sin(k_y y)\cos(k_z z)\sin(\omega t - \varphi_{y0}) \hat{\boldsymbol{y}}$$
$$+ |H_{z_0}| \cos(k_y y)\sin(k_z z)\sin(\omega t - \varphi_{z0}) \hat{\boldsymbol{z}} \big]. \tag{14}$$

$$\boldsymbol{J}_s(x = L_x, y, z, t) = 8\cos(\pi n_x) \big[ |H_{z_0}| \cos(k_y y)\sin(k_z z)\sin(\omega t - \varphi_{z0}) \hat{\boldsymbol{y}}$$
$$- |H_{y_0}| \sin(k_y y)\cos(k_z z)\sin(\omega t - \varphi_{y0}) \hat{\boldsymbol{z}} \big]. \tag{15}$$

The Lorentz force-density (i.e., force per unit area) due to the action of the $B$-field on the above surface current is thus seen to be

$$\boldsymbol{F} = \tfrac{1}{2} \boldsymbol{J}_s \times \mu_0 \boldsymbol{H} = 32\mu_0 \big[ |H_{y_0}|^2 \sin^2(k_y y)\cos^2(k_z z)\sin^2(\omega t - \varphi_{y0})$$
$$+ |H_{z_0}|^2 \cos^2(k_y y)\sin^2(k_z z)\sin^2(\omega t - \varphi_{z0}) \big] \hat{\boldsymbol{x}}. \tag{16}$$

Once again, the factor ½ in Eq.(16) arises from the fact that the effective $B$-field acting on the surface current is only half the $B$-field at the mirror surface, as the $B$-field inside the perfectly conducting wall is zero. Time averaging as well as averaging over the $y$ and $z$ coordinates yields $\langle F_x \rangle = 4\mu_0 (|H_{y0}|^2 + |H_{z0}|^2)$. Next, we sum the average force densities exerted on the three walls of the box that are perpendicular to the $x$, $y$, and $z$ axes to obtain

$$\langle F_x \rangle + \langle F_y \rangle + \langle F_z \rangle = -4\varepsilon_0 \big(|E_{x0}|^2 + |E_{y0}|^2 + |E_{z0}|^2\big) + 8\mu_0 \big(|H_{x0}|^2 + |H_{y0}|^2 + |H_{z0}|^2\big)$$
$$= 4\varepsilon_0 \big(|E_{x0}|^2 + |E_{y0}|^2 + |E_{z0}|^2\big). \tag{17}$$

Here we have used the fact that, for plane waves, $\varepsilon_0 \boldsymbol{E}_0 \cdot \boldsymbol{E}_0 = \mu_0 \boldsymbol{H}_0 \cdot \boldsymbol{H}_0$ (see Appendix B). The right-hand-side of Eq.(17) is the average energy-density of the mode, namely,

$$\langle \tfrac{1}{2}\varepsilon_0 \boldsymbol{E} \cdot \boldsymbol{E} + \tfrac{1}{2}\mu_0 \boldsymbol{H} \cdot \boldsymbol{H} \rangle = \varepsilon_0 \langle \boldsymbol{E} \cdot \boldsymbol{E} \rangle = 4\varepsilon_0 \boldsymbol{E}_0 \cdot \boldsymbol{E}_0, \tag{18}$$

as can be readily confirmed by inspecting Eq.(2). Since Eq.(17) is valid for each and every mode, its sum over all admissible modes leads once again to Eq.(12), which relates the overall pressure $p(T)$ to the total energy-density $u(T)$. Here we have invoked the symmetry of the problem, which ensures that the pressure is the same on all three walls, thus allowing one to replace the left-hand-side of Eq.(17) with $3p(T)$, albeit after adding up the contributions of all admissible modes.

**5. Blackbody radiation emerging from a gas-filled chamber**. Returning to Fig.1, we now let the cuboid chamber of dimensions $L_x \times L_y \times L_z$ with perfectly-electrically-conducting (PEC) walls be filled with a homogeneous and transparent medium (e.g., a dilute gas) of refractive index $n(\omega)$. As before, we assume that the walls of the box at temperature $T$ are in thermal equilibrium with the EM radiation inside the box. Moreover, we assume that the transparent medium that has uniformly filled the chamber is also in thermal equilibrium with the walls. Now, the wavelength of each mode is reduced by the factor $n(\omega)$, which causes the mode-density inside the chamber to be that of an empty chamber multiplied by $n^3(\omega)$. The spectral energy-density $\mathcal{E}(\omega,T)$ of the trapped radiation is thus greater, by the cube of the refractive index, than that in an otherwise empty chamber.

Despite the increased intensity of trapped radiation, the Stefan-Boltzmann law will not be affected by the presence of the medium that occupies the chamber. This is because the rays attempting to escape the cavity through the small aperture depicted in Fig.3 will suffer total internal reflection when their propagation angle relative to the $x$-axis exceeds the critical angle $\theta_{\text{crit}} = \sin^{-1}[1/n(\omega)]$. (In the following discussion we shall ignore the Fresnel reflection losses at the interface between the host medium of refractive index $n(\omega)$ and the free-space region outside the box.) For a ray propagating inside the box at an angle $\theta$ relative to the $x$-axis, the volume of a small parallelepiped with base $\delta A$ and height $(c/n)\delta t$ along the incident ray will be $(c/n)\delta t \delta A \cos\theta$. As mentioned earlier, half the EM energy inside this volume will exit through the aperture during the time $\delta t$ provided that $\theta < \theta_{\text{crit}}$. Now, the average value of $\cos\theta$ for rays in the first octant of the $k_x k_y k_z$-space that leave the chamber is given by

$$\int_0^{\theta_{\text{crit}}} \cos\theta \sin\theta \, d\theta = \tfrac{1}{2}\sin^2\theta_{\text{crit}} = 1/(2n^2), \tag{19}$$

which is the value of ½ obtained earlier for an empty chamber divided by $n^2(\omega)$. Another division by $n(\omega)$ is occasioned by the aforementioned fact that the speed of light inside the chamber has dropped from $c$ to $c/n(\omega)$. The net result is that the emitted energy (per unit area per unit time per unit angular frequency) must now be divided by $n^3(\omega)$, which cancels out the corresponding increase in the mode-density within the chamber. The Stefan-Boltzmann emission law given by Eqs.(10) and (11) thus remains unchanged, in spite of the $n^3(\omega)$ enhancement of the EM energy-density within the material medium that has uniformly filled the chamber.

**6. Electromagnetic momentum inside a Fabry-Perot cavity**. Figure 4 shows a Fabry-Perot cavity formed between two perfectly-electrically-conducting (PEC) flat mirrors located at $z = 0$ and $z = L$. The medium filling the gap between the mirrors has a refractive index $n$, and the EM wave trapped between the mirrors is the superposition of a pair of plane-waves propagating along the $z$-axis, each having angular frequency $\omega$, $k$-vector $\boldsymbol{k} = \pm(n\omega/c)\hat{\boldsymbol{z}}$, and wavelength $\lambda = 2\pi c/(n\omega)$. The gap between the mirrors is an integer-multiple of half the wavelength, i.e., $L = m\lambda/2 = m\pi c/(n\omega)$, where $m = 1, 2, 3, \cdots$ is an arbitrary positive integer. Assuming the $E$-field is linearly polarized along the $x$-axis, the $E$- and $H$-field distributions within the cavity are

$$\boldsymbol{E}(z, t) = E_0[\cos(kz - \omega t) - \cos(kz + \omega t)]\hat{\boldsymbol{x}} = 2E_0 \sin(kz)\sin(\omega t)\hat{\boldsymbol{x}}. \tag{20a}$$

$$\boldsymbol{H}(z, t) = H_0[\cos(kz - \omega t) + \cos(kz + \omega t)]\hat{\boldsymbol{y}} = 2(nE_0/Z_0)\cos(kz)\cos(\omega t)\hat{\boldsymbol{y}}. \tag{20b}$$

Here $Z_0 = \sqrt{\mu_0/\varepsilon_0}$ is the impedance of free space, while $c = 1/\sqrt{\mu_0\varepsilon_0}$ is the speed of light in vacuum. In the *SI* system of units adopted here, the permeability and permittivity of free space are denoted by $\mu_0$ and $\varepsilon_0$, respectively. It is further assumed that the gap medium is non-magnetic, that is, its relative permeability is $\mu(\omega) = 1.0$, while its relative permittivity is given by $\varepsilon(\omega) = n^2$. The dielectric susceptibility of the gap medium is thus specified as $\chi(\omega) = n^2 - 1$.

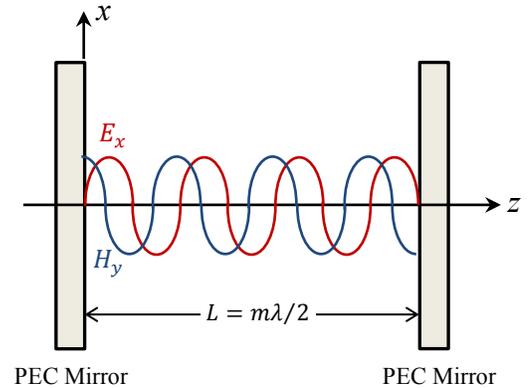

**Fig.4**. Fabry-Perot cavity formed between a pair of parallel PEC mirrors. The gap medium has length $L$ and refractive index $n$. The standing wave residing in the region between the mirrors has angular frequency $\omega$, wavelength $\lambda = 2\pi c/(n\omega)$, electric field $E_x$, and magnetic field $H_y$.

The energy-density of the EM field may now be calculated as follows [2,11]:

$$\mathcal{E}(z, t) = \tfrac{1}{2}\varepsilon_0 \varepsilon \boldsymbol{E} \cdot \boldsymbol{E} + \tfrac{1}{2}\mu_0 \boldsymbol{H} \cdot \boldsymbol{H}$$

$$= 2\varepsilon_0 n^2 E_0^2 \sin^2(kz)\sin^2(\omega t) + 2\mu_0 H_0^2 \cos^2(kz)\cos^2(\omega t)$$

$$= n^2 \varepsilon_0 E_0^2 \{[1 - \cos(2kz)]\sin^2(\omega t) + [1 + \cos(2kz)]\cos^2(\omega t)\}$$

$$= n^2 \varepsilon_0 E_0^2 [1 + \cos(2kz)\cos(2\omega t)]. \tag{21}$$
7

Denoting the cross-sectional area of the cavity by $A$, the total EM energy trapped in the cavity, $n^2\varepsilon_0 E_0^2 AL$, is seen to be constant in time. One may argue that, at any given instant, the number of photons propagating rightward (and also leftward) is $\tfrac{1}{2}n^2\varepsilon_0 E_0^2 AL/\hbar\omega$. Considering that the propagation velocity is $c/n$, the average number of photons per unit area per unit time that impinge on each mirror is seen to be $\tfrac{1}{2}n\varepsilon_0 E_0^2 c/\hbar\omega$.

The radiation pressure (i.e., force per unit area) on the PEC mirror located at $z = L$ is evaluated in accordance with the Lorentz force law [2,7,11], as follows:

$$\boldsymbol{F}(z=L,t) = \tfrac{1}{2}\boldsymbol{J}_s \times \boldsymbol{B} = (nE_0/Z_0)\cos(kL)\cos(\omega t)\hat{\boldsymbol{x}} \times 2\mu_0(nE_0/Z_0)\cos(kL)\cos(\omega t)\hat{\boldsymbol{y}}$$
$$= 2n^2\varepsilon_0 E_0^2 \cos^2(\omega t)\hat{\boldsymbol{z}}. \tag{22}$$

In the above equation, $\boldsymbol{J}_s = \boldsymbol{H}(z=L^-,t)\times\hat{\boldsymbol{z}}$ is the surface-current-density at the mirror surface, and $\boldsymbol{B} = \mu_0\boldsymbol{H}(z=L^-,t)$ is the magnetic $B$-field immediately before the mirror. The factor $\tfrac{1}{2}$ appearing in the Lorentz force expression accounts for the fact that the effective $B$-field acting on the surface-current $\boldsymbol{J}_s$ is the average of the $B$-fields at $z=L^+$, where $\boldsymbol{B}=0$, and $z=L^-$. (The pressure on the mirror at $z=0$ is the same as that given by Eq.(22), except that its direction is reversed.) Note that the number of photons impinging on each mirror per unit area per unit time, which we obtained earlier, must be multiplied by $2n\hbar\omega/c$ in order to arrive at the time-averaged pressure $\langle F_z(z=L,t)\rangle = n^2\varepsilon_0 E_0^2$ given by Eq.(22). This fact has been interpreted by some authors in the past as supporting evidence that each photon residing in a dielectric medium of refractive index $n$ carries the Minkowski momentum $n\hbar\omega/c$ [3-6]. In what follows, we shall argue that the radiation pressure exerted on the mirrors in accordance with Eq.(22) is consistent with a nuanced interpretation of Abraham's photon momentum [7,8].

Let us then compute the Abraham momentum-density $\boldsymbol{\mathcal{P}}_A(z,t) = \boldsymbol{E}\times\boldsymbol{H}/c^2$ of the EM field that forms the standing wave inside the dielectric-filled cavity. Recall that, for a single plane-wave propagating in a dielectric host of refractive index $n$, the Poynting vector $\boldsymbol{S}(\boldsymbol{r},t) = \boldsymbol{E}\times\boldsymbol{H}$ yields the EM energy that crosses a unit area per unit time [2,11]. The volume occupied by this energy is $c/n$, which also contains a total Abraham momentum of $\boldsymbol{E}\times\boldsymbol{H}/(nc)$. Normalizing this EM momentum by the corresponding EM energy, and recalling that individual photons have energy $\hbar\omega$, one concludes that the Abraham momentum per photon is $\hbar\omega/(nc)$. In the case of the standing wave specified by Eqs.(20a) and (20b), we find the EM momentum-density to be

$$\boldsymbol{\mathcal{P}}_A(z,t) = (n\varepsilon_0 E_0^2/c)\sin(2kz)\sin(2\omega t)\hat{\boldsymbol{z}}. \tag{23}$$

The dielectric slab of thickness $L = m\lambda/2$ thus appears as a contiguous series of $2m$ layers of thickness $\lambda/4$ demarcated by the adjacent zeros of $\sin(2kz)$, similar to the one between $z = (2m-1)\lambda/4$ and $z = m\lambda/2 = L$. The EM (Abraham) momentum per unit cross-sectional area contained in each such layer is readily determined as follows:

$$\int_{(2m-1)\lambda/4}^{m\lambda/2}\boldsymbol{\mathcal{P}}_A(z,t)\mathrm{d}z = -(\varepsilon_0 E_0^2/\omega)\sin(2\omega t)\hat{\boldsymbol{z}}. \tag{24}$$

Note that the EM momentum content of each layer oscillates between $\pm(\varepsilon_0 E_0^2/\omega)\hat{\boldsymbol{z}}$ with frequency $2\omega$, even though, in accordance with Eq.(21), the EM energy content of each layer is fixed at $n^2\varepsilon_0 E_0^2\lambda/4 = \pi n E_0^2/(2Z_0\omega)$. The time-rate-of-change of the momentum content of the layer may now be determined by differentiating Eq.(24), as follows:

$$\frac{\mathrm{d}}{\mathrm{d}t}\int_{(2m-1)\lambda/4}^{m\lambda/2}\boldsymbol{\mathcal{P}}_A(z,t)\mathrm{d}z = -2\varepsilon_0 E_0^2\cos(2\omega t)\hat{\boldsymbol{z}}. \tag{25}$$

Before using the above result, we need an expression for the time-rate-of-change of the *mechanical* momentum residing in each layer, which is the EM force exerted on the electric dipoles that constitute the dielectric material. The EM force-density acting on the polarization $\boldsymbol{P}(\boldsymbol{r},t) = \varepsilon_0(\varepsilon-1)\boldsymbol{E}(\boldsymbol{r},t)$ of the dielectric medium within the Fabry-Perot cavity is given by [3-9]

$$\boldsymbol{F}(z,t) = (\boldsymbol{P} \cdot \boldsymbol{\nabla})\boldsymbol{E}^{\nearrow 0} + (\partial \boldsymbol{P}/\partial t) \times \mu_0 \boldsymbol{H}$$

$$= 2\varepsilon_0(\varepsilon - 1)\omega E_0 \sin(kz)\cos(\omega t)\hat{\boldsymbol{x}} \times 2\mu_0(nE_0/Z_0)\cos(kz)\cos(\omega t)\hat{\boldsymbol{y}}$$

$$= 2(n^2 - 1)\varepsilon_0 E_0^2 (n\omega/c) \sin(2kz) \cos^2(\omega t)\hat{\boldsymbol{z}}. \tag{26}$$

Integrating the above force-density over the thickness of a $\lambda/4$-thick layer, we find

$$\int_{(2m-1)\lambda/4}^{m\lambda/2} \boldsymbol{F}(z,t)\mathrm{d}z = -2(n^2 - 1)\varepsilon_0 E_0^2 \cos^2(\omega t)\hat{\boldsymbol{z}}. \tag{27}$$

The time-averaged EM force per unit cross-sectional area of each $\lambda/4$-thick layer is given by $\pm(n^2 - 1)\varepsilon_0 E_0^2 \hat{\boldsymbol{z}}$. Adjacent layers, of course, experience equal but opposite forces and, given that the total number $2m$ of such layers is even, the overall EM force experienced by the dielectric slab of thickness $L$ is seen to be zero at all times. Focusing our attention on the $\lambda/4$-thick dielectric layer that is in contact with the mirror at $z = L$, we conclude that the time-rate-of-change of the *total* momentum content of this layer is the sum of contributions from the EM momentum, as given by Eq.(25), and from the mechanical momentum, as given by Eq.(27).

In order to tie together the various results of the preceding discussion, we need to bring in one final ingredient of the classical theory of electrodynamics, namely, the EM stress tensor, which provides the rate of flow of the EM momentum per unit time per unit cross-sectional area along the $x$, $y$, and $z$ directions [2,7]. The Einstein-Laub stress tensor of the EM field is given by [8]

$$\overleftrightarrow{\mathcal{T}}(\boldsymbol{r},t) = \tfrac{1}{2}(\varepsilon_0 \boldsymbol{E} \cdot \boldsymbol{E} + \mu_0 \boldsymbol{H} \cdot \boldsymbol{H})\overleftrightarrow{\mathbf{I}} - \boldsymbol{D}\boldsymbol{E} - \boldsymbol{B}\boldsymbol{H}. \tag{28}$$

Consequently, for the standing wave described by Eqs.(20a) and (20b), the rate of flow of EM linear momentum along the $z$-axis is determined as follows:

$$\mathcal{T}_{zz}(z,t) = \tfrac{1}{2}\varepsilon_0 \boldsymbol{E} \cdot \boldsymbol{E} + \tfrac{1}{2}\mu_0 \boldsymbol{H} \cdot \boldsymbol{H}$$

$$= 2\varepsilon_0 E_0^2 \sin^2(kz)\sin^2(\omega t) + 2n^2 \varepsilon_0 E_0^2 \cos^2(kz)\cos^2(\omega t)$$

$$= \varepsilon_0 E_0^2 [1 - \cos(2kz)]\sin^2(\omega t) + n^2 \varepsilon_0 E_0^2 [1 + \cos(2kz)]\cos^2(\omega t)$$

$$= \varepsilon_0 E_0^2 \{[1 + (n^2 - 1)\cos^2(\omega t)] - [1 - (n^2 + 1)\cos^2(\omega t)]\cos(2kz)\}. \tag{29}$$

At $z = L$, where $\cos(2kz) = \cos(2m\pi) = 1$, the outflow of the EM momentum per unit cross-sectional area is $2n^2 \varepsilon_0 E_0^2 \cos^2(\omega t)$, which is consistent with the pressure exerted on the mirror in accordance with Eq.(22). At $z = L - \tfrac{1}{4}\lambda$, however, $\cos(2kz) = -1$, and the rate of flow of EM momentum into the $\lambda/4$-thick layer that is in contact with the mirror is $2\varepsilon_0 E_0^2 \sin^2(\omega t)$. From this we must now subtract the time-rate-of-change of the *total* momentum content of the dielectric layer, i.e., the sum of the expressions given by Eqs.(25) and (27). We will have

$$\mathcal{T}_{zz}(L - \tfrac{1}{4}\lambda, t)\hat{\boldsymbol{z}} - \tfrac{\mathrm{d}}{\mathrm{d}t}\int_{L - \tfrac{1}{4}\lambda}^{L} \boldsymbol{\mathcal{P}}_A(z,t)\mathrm{d}z - \int_{L - \tfrac{1}{4}\lambda}^{L} \boldsymbol{F}(z,t)\mathrm{d}z$$

$$= 2\varepsilon_0 E_0^2 [\sin^2(\omega t) + \cos(2\omega t) + (n^2 - 1)\cos^2(\omega t)]\hat{\boldsymbol{z}}$$

$$= 2n^2 \varepsilon_0 E_0^2 \cos^2(\omega t)\hat{\boldsymbol{z}}. \tag{30}$$

The final expression in Eq.(30) is the rate of outflow of EM momentum at $z = L$, which is also the pressure exerted on the mirror at $z = L$; see Eq.(22). We conclude that the force exerted on the end mirrors is consistent with the Abraham momentum-density of the EM field trapped within the Fabry-Perot cavity, as given by Eq.(23). In the process, the field also exerts a force on the dielectric slab, with the force-density given by Eq.(26). The argument presented here is thus seen to be consistent with the assertion that individual photons propagating along the $+z$ and $-z$ directions inside the cavity possess the EM Abraham momentum $\pm(\hbar\omega/nc)\hat{\boldsymbol{z}}$.

**7. Energy-density fluctuations**. The square of the photon number $m$ contained in a given radiation mode trapped inside the cavity of Fig.1 is readily found to be

$$\langle m^2 \rangle = [1 - \exp(-\hbar\omega/k_BT)] \sum_{m=0}^{\infty} m^2 \exp(-m\hbar\omega/k_BT)$$

$$= [1 - \exp(-\hbar\omega/k_BT)](k_BT/\hbar)^2 \frac{d^2}{d\omega^2} \sum_{m=0}^{\infty} \exp(-m\hbar\omega/k_BT)$$

$$= [1 - \exp(-\hbar\omega/k_BT)](k_BT/\hbar)^2 \frac{d^2}{d\omega^2} [1 - \exp(-\hbar\omega/k_BT)]^{-1}$$

$$= \frac{\exp(\hbar\omega/k_BT) + 1}{[\exp(\hbar\omega/k_BT) - 1]^2}. \tag{31}$$

Consequently, the photon number variance is

$$\langle m^2 \rangle - \langle m \rangle^2 = \frac{\exp(\hbar\omega/k_BT)}{[\exp(\hbar\omega/k_BT) - 1]^2}. \tag{32}$$

The variance of energy per unit volume per unit angular frequency is thus found to be

$$\text{Var}[\mathcal{E}(\omega,T)] = \frac{\hbar^2\omega^4 \exp(\hbar\omega/k_BT)}{\pi^2 c^3 [\exp(\hbar\omega/k_BT) - 1]^2} = \hbar\omega\mathcal{E}(\omega,T) + \pi^2 c^3 [\mathcal{E}(\omega,T)/\omega]^2. \tag{33}$$

(As an aside, note that the second term on the right-hand-side of Eq.(33) is the classical expression of the EM field-intensity fluctuations, which must be approached when $\hbar \to 0$.) If we now define the average-to-standard-deviation ratio for the energy-density of thermal radiation, Eqs.(8) and (33) yield

$$\frac{\text{Average } \mathcal{E}(\omega,T)}{\text{Standard Deviation}} = \frac{\omega \exp(-\hbar\omega/2k_BT)}{\pi c^{3/2}}. \tag{34}$$

Note that the above ratio has the dimensions of $\sqrt{s/m^3}$, which indicates that the right-hand-side of Eq.(34) must be multiplied by $\sqrt{vd\omega}$ in order to yield the average-to-standard-deviation ratio for the radiation contained in a volume $v$ within the frequency range $d\omega$. Considering that the various radiation modes are independent of each other, the total variance (per unit volume) can be obtained by integrating Eq.(33) over all frequencies, as follows:

$$\int_0^\infty \text{Var}[\mathcal{E}(\omega,T)] d\omega = \int_0^\infty \frac{\hbar^2 \omega^4 \exp(\hbar\omega/k_BT)}{\pi^2 c^3 [\exp(\hbar\omega/k_BT) - 1]^2} d\omega$$

$$= \frac{k_B^5 T^5}{\pi^2 c^3 \hbar^3} \int_0^\infty \frac{x^4 \exp(x)}{[\exp(x) - 1]^2} dx = -\frac{k_B^5 T^5}{\pi^2 c^3 \hbar^3} \int_0^\infty x^4 \frac{d}{dx} [\exp(x) - 1]^{-1} dx$$

$$\xrightarrow{\text{Integration by parts}} = \frac{4k_B^5 T^5}{\pi^2 c^3 \hbar^3} \int_0^\infty \frac{x^3}{\exp(x) - 1} dx \xrightarrow{\text{G\&R 3.411-17 [26]}} \frac{4\pi^2 k_B^5 T^5}{15 c^3 \hbar^3} = 16(k_B\sigma/c)T^5. \tag{35}$$

The above result is in agreement with the thermodynamic identity $\langle \varepsilon^2 \rangle - \langle \varepsilon \rangle^2 = k_B T^2 \partial\langle\varepsilon\rangle/\partial T$. Given that $\sigma$ has the dimensions of $J/(s \cdot m^2)$, the coefficient of $T^5$ in the preceding equation is seen to have the dimensions of $J^2/m^3$. Dividing the average energy-density, $(4\sigma/c)T^4$, by the square root of the variance given by Eq.(35), we find the average-to-standard-deviation ratio for the energy of thermal radiation contained in a volume $V$ to be $\sqrt{(\sigma V/ck_B)T^3}$.

As for the total energy (per unit area per unit time) emerging from an aperture on the surface of the blackbody depicted in Fig.3, the average is $\sigma T^4$ while the variance is $4k_B \sigma T^5$, obtained by multiplying the expression in Eq.(35) by $c/4$. The average-to-standard-deviation ratio for the EM energy emerging from an aperture of area $A$ during a time interval $\tau$ is thus $\sqrt{(\sigma A\tau/4k_B)T^3}$.

**8. Entropy of thermal radiation.** Let an evacuated chamber of volume $V$ contain a photon gas in thermal equilibrium with the walls of the chamber at temperature $T$. Each EM mode could have $m$ photons ($m = 0, 1, 2, \cdots$) with probability $p_m = [1 - \exp(-\hbar\omega/k_B T)]\exp(-m\hbar\omega/k_B T)$. The entropy $S_{\text{mode}}$ of each such mode is, therefore, given by

$$\begin{aligned}
S_{\text{mode}}(\omega, T) &= -k_B \sum_{m=0}^{\infty} p_m \ln p_m \\
&= -k_B \sum_{m=0}^{\infty} p_m \{\ln[1 - \exp(-\hbar\omega/k_B T)] - (m\hbar\omega/k_B T)\} \\
&= (\hbar\omega/T)\langle m \rangle - k_B \ln[1 - \exp(-\hbar\omega/k_B T)] \\
&= \frac{\hbar\omega/T}{\exp(\hbar\omega/k_B T) - 1} - k_B \ln[1 - \exp(-\hbar\omega/k_B T)].
\end{aligned} \qquad (36)$$

Since the various modes are independent of each other, the total entropy of the trapped photon gas is the sum of the entropies of all available modes. Given that the mode density (per unit volume) is $\omega^2 d\omega/(\pi^2 c^3)$, the entropy of the photon gas is readily evaluated by integrating over all frequencies $\omega$, as follows:

$$\begin{aligned}
S(T, V) &= \int_0^\infty \left(\frac{V\omega^2}{\pi^2 c^3}\right) S_{\text{mode}}(\omega, T) d\omega \\
&= \left(\frac{k_B V}{\pi^2 c^3}\right) \int_0^\infty \left\{\frac{\hbar\omega^3/k_B T}{\exp(\hbar\omega/k_B T) - 1} - \underbrace{\omega^2 \ln[1 - \exp(-\hbar\omega/k_B T)]}_{\text{Use integration by parts}}\right\} d\omega \\
&= \left(\frac{4k_B V}{3\pi^2 c^3}\right) \int_0^\infty \frac{\hbar\omega^3/k_B T}{\exp(\hbar\omega/k_B T) - 1} d\omega \\
&= \left(\frac{4k_B V}{3\pi^2 c^3}\right) \left(\frac{k_B T}{\hbar}\right)^3 \underbrace{\int_0^\infty \frac{x^3 dx}{\exp(x) - 1}}_{\pi^4/15} = \left(\frac{4\pi^2 k_B^4}{45 c^3 \hbar^3}\right) V T^3.
\end{aligned} \qquad (37)$$

An alternative derivation of the above result assumes that an evacuated chamber of fixed volume $V$ is slowly heated from 0K to the temperature $T$. The entropy of the photon gas is then evaluated by integrating $\Delta Q/T$ with respect to the rising temperature from 0 to $T$. Since the heat flows into the chamber in increments of $\Delta Q = V\Delta u = [\pi^2 k_B^4 V/(15 c^3 \hbar^3)]\Delta(T^4)$, we will have

$$S(T, V) = \int_0^T \frac{dQ}{T} = \left(\frac{\pi^2 k_B^4 V}{15 c^3 \hbar^3}\right) \int_0^T \frac{d(T^4)}{T} = \left(\frac{4\pi^2 k_B^4 V}{15 c^3 \hbar^3}\right) \int_0^T T^2 dT = \left(\frac{4\pi^2 k_B^4}{45 c^3 \hbar^3}\right) V T^3. \qquad (38)$$

Alternatively, one may imagine an evacuated chamber at a constant temperature $T$, whose volume slowly (i.e., reversibly) rises from $V_1$ to $V_2$. Considering that $\Delta Q = \Delta U + P\Delta V$, and that the radiation pressure inside the chamber is $P = u(T)/3$ while $u(T) = \pi^2 k_B^4 T^4/(15 c^3 \hbar^3)$, we will have

$$\begin{aligned}
S(T, V_2) - S(T, V_1) &= \int_{V_1}^{V_2} \frac{dQ}{T} = \int_{V_1}^{V_2} \frac{u(T) + \tfrac{1}{3}u(T)}{T} dV \\
&= \left(\frac{4\pi^2 k_B^4}{45 c^3 \hbar^3}\right) \int_{V_1}^{V_2} T^3 dV = \left(\frac{4\pi^2 k_B^4 T^3}{45 c^3 \hbar^3}\right) (V_2 - V_1).
\end{aligned} \qquad (39)$$

Clearly, all these different approaches to computing the entropy of the photon gas in thermal equilibrium with its container of volume $V$ and temperature $T$ yield identical results. We note in passing that the energy fluctuations of thermal radiation derived earlier (see Eq.(35)) could also be obtained from either of the thermodynamic identities $\langle \varepsilon^2 \rangle - \langle \varepsilon \rangle^2 = k_B T^3 (\partial S/\partial T)$ or $\langle \varepsilon^2 \rangle - \langle \varepsilon \rangle^2 = -k_B (\partial^2 S/\partial \varepsilon^2)^{-1}$.

**9. Dielectric particle in thermal equilibrium with electromagnetic radiation**. This section is devoted to an alternative derivation of Planck's blackbody radiation formula, Eq.(8). Let a small spherical particle of volume $v$ and dielectric susceptibility $\varepsilon_0 \chi(\omega)$ — when subjected to an external $E$-field $E_0 \hat{x} \cos(\omega t)$ — be in thermal equilibrium with the radiation inside an otherwise empty box at temperature $T$. The Lorentz oscillator model for the particle yields the following formula for its susceptibility at frequency $\omega$:

$$\chi(\omega) = \frac{\omega_p^2}{\omega_0^2 - \omega^2 - i[\gamma\omega + (4\pi^2 \omega_p^2/3)(v/\lambda_0^3)]}. \tag{40}$$

In the above equation, $\omega_p^2 = Nq^2/(m\varepsilon_0)$ is the plasma frequency and $\omega_0 = \sqrt{\alpha/m}$ the resonance frequency of the material, $\gamma = \beta/m$ is the damping coefficient associated with loss mechanisms other than radiation resistance (e.g., absorption losses), and $\lambda_0 = 2\pi c/\omega$ is the vacuum wavelength of the excitation field. Here $N$ is the number-density of the electric dipoles that comprise the spherical particle, $q$ and $m$ are the electric charge and mass of the oscillating electron within individual dipoles, $\alpha$ is the dipole's spring constant, $\beta$ is the dipole's friction coefficient, and $\varepsilon_0$ is the permittivity of free space.

In what follows, we shall assume that the particle is transparent and that, therefore, $\gamma$ can be set to zero. The only damping mechanism will then be radiation resistance, whose contribution appears as the second term inside the square brackets in Eq.(40). If we further assume that the particle has a narrow resonance peak at $\omega = \omega_0$, we can define the effective damping coefficient $\gamma_{\text{eff}} = v\omega_p^2 \omega_0^2/(6\pi c^3) = \mu_0 \tilde{q}^2 \omega_0^2/(6\pi \tilde{m} c)$, where $\tilde{q}$ is the *total* charge and $\tilde{m}$ the *total* mass of oscillating electrons.

At thermal equilibrium, each vibrational degree of freedom of the spherical particle has, on average, $\tfrac{1}{2} k_B T$ of energy. Given that harmonic oscillators have two degrees of freedom (associated with their kinetic and potential energies), the oscillators along the $x$, $y$, and $z$ axes will each have an average thermal energy of $k_B T$, for a total thermal energy of $3 k_B T$. Now, the impulse response of the Lorentz oscillator diminishes over time as $\exp(-\gamma t/2)$, indicating that the radiated energy decays as $\exp(-\gamma t)$. All in all, one can say that, in thermal equilibrium, the randomly vibrating particle radiates EM energy at an average rate of $3\gamma_{\text{eff}} k_B T$ Joule/sec [11].

The oscillators are immersed in a bath of EM radiation. Let the $E$-field acting on the particle be $\boldsymbol{E}(t) = E_0(\omega)\hat{x} \exp(-i\omega t)$. Then the rate at which the particle extracts energy from the field will be

$$\langle \boldsymbol{E}(t) \cdot d\boldsymbol{p}(t)/dt \rangle = \tfrac{1}{2}\text{Re}[-i\omega \varepsilon_0 \chi(\omega) v |E_0(\omega)|^2] = \tfrac{1}{2}\left[\frac{\gamma_{\text{eff}} \omega_p^2 v \omega^2}{(\omega_0^2 - \omega^2)^2 + \gamma_{\text{eff}}^2 \omega^2}\right]\varepsilon_0 |E_0(\omega)|^2. \tag{41}$$

Considering the sharpness of the resonance line around the center-frequency $\omega = \omega_0$, we may approximate the term $\omega_0^2 - \omega^2$ with $(\omega_0 + \omega)(\omega_0 - \omega) \cong 2\omega_0(\omega_0 - \omega)$, and $\tfrac{1}{2}\varepsilon_0|E_0(\omega)|^2$ with $\tfrac{1}{2}\varepsilon_0|E_0(\omega_0)|^2 = \mathcal{E}(\omega_0, T)$, which is the total (i.e., electric plus magnetic) field energy-density (i.e., energy per unit volume per unit frequency) at the resonance frequency of the oscillator at temperature $T$. Integrating over $\omega$ (in order to cover the entire linewidth) now yields

$$\int_0^\infty \langle \boldsymbol{E}(t) \cdot d\boldsymbol{p}(t)/dt \rangle d\omega \cong \mathcal{E}(\omega_0, T) \int_{-\infty}^\infty \frac{\gamma_{\text{eff}} \omega_p^2 v}{4(\omega - \omega_0)^2 + \gamma_{\text{eff}}^2} d\omega = \tfrac{1}{2}\pi \omega_p^2 v \mathcal{E}(\omega_0, T). \tag{42}$$

At thermal equilibrium, where the energy radiation and absorption rates computed above are equal, we will have

$$\tfrac{1}{2}\pi \omega_p^2 v \mathcal{E}(\omega_0, T) = 3\gamma_{\text{eff}} k_B T \quad \rightarrow \quad \mathcal{E}(\omega, T) = k_B T \omega^2/(\pi^2 c^3). \tag{43}$$

The above equation, the well-known Rayleigh-Jeans formula for the EM energy-density inside a black box at temperature $T$, is known to be wrong because it predicts, unrealistically, the existence of high EM energy-densities at high frequencies (in proportion to $\omega^2$). This failure of classical physics at the end of the 19th century came to be known as the *ultraviolet catastrophe*. The ultraviolet catastrophe was eventually resolved when Max Planck proposed his quantum hypothesis, which allows a (harmonic) oscillator to possess only energies that are integer-multiples of $\hbar \omega_0$. Here $\hbar$ is the reduced Planck constant while $\omega_0$ is the resonance frequency of the harmonic oscillator. The oscillator will thus have an energy $n\hbar\omega_0$ with the probability $p_n = \alpha \exp(-n\hbar\omega_0/k_B T)$; here $n = 0, 1, 2, 3, \cdots$ is the integer-valued quantum number that specifies the allowed energy levels of the oscillator. The coefficient $\alpha$ is determined from the fact that the sum of the probabilities must equal 1.0, that is,

$$\sum_{n=0}^{\infty} p_n = \alpha \sum_{n=0}^{\infty} \exp(-n\hbar\omega_0/k_B T) = \frac{\alpha}{1 - \exp(-\hbar\omega_0/k_B T)} = 1$$

$$\rightarrow \quad \alpha = 1 - \exp(-\hbar\omega_0/k_B T). \tag{44}$$

Consequently, the average energy of the harmonic oscillator is found to be

$$\sum_{n=0}^{\infty}(n\hbar\omega_0)p_n = \alpha\hbar\omega_0 \sum_{n=0}^{\infty} n \exp(-n\hbar\omega_0/k_B T)$$

$$= \frac{\alpha\hbar\omega_0 \exp(-\hbar\omega_0/k_B T)}{[1 - \exp(-\hbar\omega_0/k_B T)]^2} = \frac{\hbar\omega_0}{\exp(\hbar\omega_0/k_B T) - 1}. \tag{45}$$

Only in the limit of large $T$ and/or small $\omega_0$ does the average energy (kinetic plus potential) of the oscillator given by Eq.(45) approach the classical value of $k_B T$. The correct formula for the energy-density of blackbody radiation is thus obtained by replacing $k_B T$ in Eq.(43) with the average energy of the oscillator given by Eq.(45). We will have

$$\mathcal{E}(\omega, T) = \frac{\hbar\omega^3}{\pi^2 c^3 [\exp(\hbar\omega/k_B T) - 1]}. \tag{46}$$

This is the same equation, Eq.(8), derived earlier when we invoked the quantum hypothesis in conjunction with the various modes of the EM field trapped inside an empty box in thermal equilibrium with the walls at temperature $T$.

**10. Einstein's *A* and *B* coefficients**. In his famous 1916 and 1917 papers [15-18], Einstein suggested that an atom/molecule in a state $|1\rangle$, when immersed in EM radiation having local energy-density $\mathcal{E}(\omega, T)$ at frequency $\omega$ and temperature $T$, will transition to the excited state $|2\rangle$ by absorbing a quantum of energy $\hbar\omega$ from the field at a rate proportional to $I(\omega, T) = c\mathcal{E}(\omega, T)$. The energy difference between $|1\rangle$ and $|2\rangle$ was noted to be $\varepsilon_2 - \varepsilon_1 = \hbar\omega$, and the proportionality coefficient for the absorption-induced transition was denoted by $B_{12}$. Einstein then suggested that the excited atom/molecule will have *two* pathways for returning to the lower-energy state: i) spontaneous emission, which occurs at a rate of, say, $A_{21}$ transitions per second, and ii) stimulated emission, whose rate is proportional to the local intensity $I(\omega, T)$, with a proportionality coefficient $B_{21}$. Given the energy difference $\Delta\varepsilon = \hbar\omega$ between $|1\rangle$ and $|2\rangle$, the probabilities of being in $|1\rangle$ and $|2\rangle$ are in the ratio of $p_2/p_1 = \exp(-\hbar\omega/k_B T)$. In thermal equilibrium, the rate of the $|1\rangle \rightarrow |2\rangle$ transition must equal that of the $|2\rangle \rightarrow |1\rangle$ transition, that is,

$$p_1 B_{12} I(\omega, T) = p_2 B_{21} I(\omega, T) + p_2 A \quad \rightarrow \quad I(\omega, T) = \frac{A}{B_{12} \exp(\hbar\omega/k_B T) - B_{21}}. \tag{47}$$

A comparison with Eq.(46) reveals that $B_{12} = B_{21}$, that is, absorption and stimulated emission have identical rates, and that $A_{21}/B_{21} = \hbar\omega^3/(\pi^2 c^2)$.

**11. Implications of the second law of thermodynamics for radiation pressure**. In this section we infer several facts about the pressure of thermal radiation using the second law. According to Feynman, "*one cannot devise a process whose only result is to convert heat to work at a single temperature*" [11]. An alternative statement of the second law, once again according to Feynman [11], is that "*heat cannot be taken in at a certain temperature and converted into work with no other change in the system or the surroundings.*" We apply the second law to a rotary machine placed within thermal radiation, because the rotary machine returns to its initial state after a full revolution. If mechanical work is performed after a full revolution of the device, and if no other changes are made in the system, then the second law will be violated.

We have seen that the pressure of thermal radiation on a perfect reflector is $p(T) = u(T)/3$, where $u(T)$ is the energy-density of the EM field at equilibrium with its container held at the temperature $T$; see

Eq.(12). Does this result change if the reflector is not flat (e.g., it has a curved surface), or if the reflector's surface is jagged (i.e., it is a scatterer)? What happens if the surface is not reflective at all, but rather an ideal absorber? What if the surface is a partial absorber (and partial reflector)?

To answer these questions, consider a rotary machine whose vanes have different optical characteristics on their opposite facets; for instance, while the front side of each vane is a flat reflector, the back side might be a hemispherical reflector, as shown in Fig.5(a). The rotor is placed inside an empty box which contains EM radiation at the equilibrium temperature $T$. If the pressure of radiation on one side of each vane happens to differ from the corresponding pressure on the opposite side, the rotor will start rotating in a fixed direction, thereby producing useful mechanical work from a single source that is maintained at a fixed temperature $T$. This will violate the second law of thermodynamics. We conclude that the net pressure of thermal radiation on the flat mirror is the same as that on the hemispherical mirror.

Figure 5(b) shows three different constructs for the vanes. Instead of hemispherical reflector, the backside of the vane may be an absorber (either perfect or partial), or it may be a scatterer. Irrespective of the nature of the back surface, the second law of thermodynamics teaches us that the pressure of radiation on that surface must always be given by $p(T) = u(T)/3$.

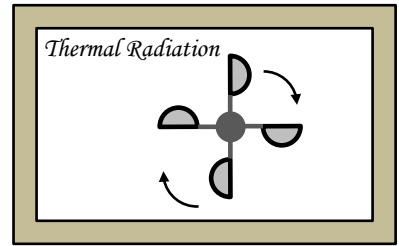

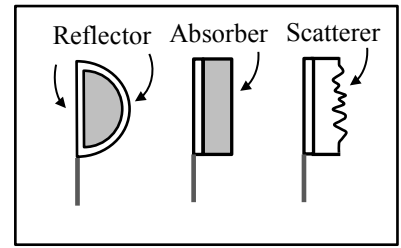

**Fig.5**. The empty box at the top of the figure is held at a constant temperature $T$. Inside the box, EM radiation is at thermal equilibrium with the enclosure. Each vane of the rotary fan consists of a solid hemisphere coated with a perfect reflector. The radiation pressure on the flat side of each vane must be equal to that on the spherical side, otherwise the fan will rotate and perform useful mechanical work, in violation of the second law of thermodynamics. The bottom figure shows three types of vane (including the hemispherical one). A vane may be perfectly reflective on one side and perfectly (or partially) absorptive on the opposite side, or it may have a reflector on one side and a scatterer on the other side. In all cases the pressure of thermal radiation must be identical on both sides of each vane.

In the case of a perfect (or partial) absorber, Kirchhoff's law affirms that, at thermal equilibrium, the absorption and emission efficiencies are identical, which explains why the pressure on the reflective side of the vane should equal that on the absorptive side. In the case of a scatterer, the reciprocity theorem of classical electrodynamics [19] must be invoked to explain the equality of the radiation pressure on the scattering and reflective sides of the vane. (The fact that all propagation directions contain, on average, equal amounts of EM energy, that they are all equally likely to exist within the cavity, and also the fact that different cavity modes are relatively incoherent, combined with the reciprocity theorem is sufficient to prove that radiation pressure on the scatterer is the same as that on a perfect reflector. In fact, one could argue that this limited form of the reciprocity theorem is an immediate consequence of the second law.)

As a second example, consider the system depicted in Fig.6(a), where the radiation emanating from a blackbody at temperature $T$ is collected by a lens and focused onto the shiny teeth of a rotating ratchet. The reflected light is Doppler shifted, as will be shown below, and the change in its energy and momentum will be transferred to the rotating ratchet. In this way the thermal radiation performs useful work without violating the second law of thermodynamics. The reason is that the reflected radiation in this case is Doppler-shifted to a lower frequency and is, therefore, at a lower temperature than the incident radiation. So long as the rotating wheel extracts thermal energy from a hot bath and returns a fraction of this energy to a cold bath, the second law would allow the remaining energy to be converted into useful (mechanical) work.

Computing the Doppler shift of the beam reflecting off the ratchet requires that we first move to the reference frame in which the illuminated tooth is stationary, as shown in Fig.6(b). A Lorentz transformation of the space-time coordinates from the $xyz$ lab frame to the $x'y'z'$ rest-frame is needed to determine the phase factor $(\boldsymbol{k} \cdot \boldsymbol{r} - \omega t)$ of the incident plane-wave, as follows:

$$\boldsymbol{k} \cdot \boldsymbol{r} - \omega t = -(\omega^{\mathrm{inc}}/c)z - \omega^{\mathrm{inc}}t = -(\omega^{\mathrm{inc}}/c)z' - \omega^{\mathrm{inc}}\gamma(t' + vx'/c^2)$$
$$= -\gamma(v/c^2)\omega^{\mathrm{inc}}x' - (\omega^{\mathrm{inc}}/c)z' - \gamma\omega^{\mathrm{inc}}t'. \tag{48}$$

This means that the incident plane-wave seen by the mirror has frequency $\omega' = \gamma\omega^{\mathrm{inc}}$ and propagates along $\boldsymbol{k}' = -\gamma(v/c^2)\omega^{\mathrm{inc}}\hat{\boldsymbol{x}} - (\omega^{\mathrm{inc}}/c)\hat{\boldsymbol{z}}$, which makes an angle $\varphi = \sin^{-1}(v/c)$ with the $z'$-axis. The incidence angle in the mirror's rest-frame is thus $\theta - \varphi$, causing the reflected $k$-vector to emerge at an angle $2\theta - \varphi$ relative to the $z'$-axis (again in the mirror's rest frame). Now, the magnitude of the $k$-vector in the rest-frame is $\omega'/c = \gamma\omega^{\mathrm{inc}}/c$ and, therefore, the reflected $k$-vector becomes

$$\boldsymbol{k}' = (\gamma\omega^{\mathrm{inc}}/c)[\sin(2\theta - \varphi)\,\hat{\boldsymbol{x}}' + \cos(2\theta - \varphi)\,\hat{\boldsymbol{z}}']. \tag{49}$$

A Lorentz transformation of the reflected plane-wave's phase-factor to the lab frame now yields

$$\boldsymbol{k}' \cdot \boldsymbol{r}' - \omega' t' = (\gamma\omega^{\mathrm{inc}}/c)[\sin(2\theta - \varphi)\,x' + \cos(2\theta - \varphi)\,z'] - \gamma\omega^{\mathrm{inc}}t'$$
$$= (\gamma\omega^{\mathrm{inc}}/c)[\sin(2\theta - \varphi)\,\gamma(x - vt) + \cos(2\theta - \varphi)\,z] - \gamma\omega^{\mathrm{inc}}\gamma(t - vx/c^2)$$
$$= \gamma^2(\omega^{\mathrm{inc}}/c)[(v/c) + \sin(2\theta - \varphi)]x + \gamma(\omega^{\mathrm{inc}}/c)\cos(2\theta - \varphi)\,z$$
$$- \gamma^2\omega^{\mathrm{inc}}[1 + (v/c)\sin(2\theta - \varphi)]t. \tag{50}$$

Thus, back in the $xyz$ lab frame, the reflected beam's frequency and wave-vector will be

$$\omega^{\mathrm{ref}} = \gamma^2\omega^{\mathrm{inc}}[1 + (v/c)\sin(2\theta - \varphi)], \tag{51a}$$
$$\boldsymbol{k}^{\mathrm{ref}} = \gamma^2(\omega^{\mathrm{inc}}/c)[(v/c) + \sin(2\theta - \varphi)]\hat{\boldsymbol{x}} + \gamma(\omega^{\mathrm{inc}}/c)\cos(2\theta - \varphi)\,\hat{\boldsymbol{z}}. \tag{51b}$$

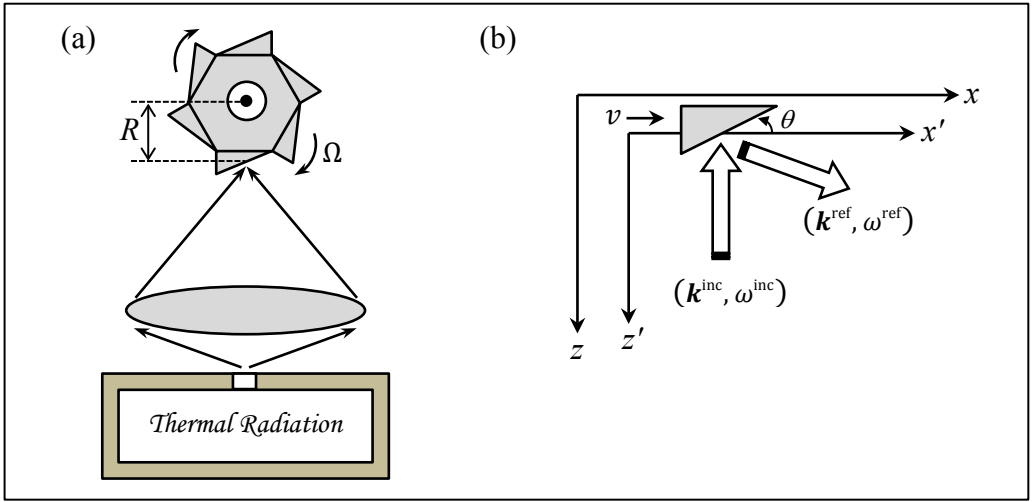

**Fig. 6**. (a) Thermal radiation from a blackbody is captured by a lens and focused onto a ratchet with reflective teeth (or facets). Each tooth of the ratchet makes an angle $\theta$ with its nominally circular boundary. The ratchet, which is rotating at a constant angular velocity $\Omega$, has average radius $R$ and moment of inertia $I$. The pressure of radiation exerts a torque on the rotating ratchet, thus producing useful work. In the process, the light reflected from the tooth is Doppler shifted, acquiring a lower temperature. This system does *not* violate the second law of thermodynamics because the reflected light essentially constitutes a cold bath. (b) A perfectly reflecting tooth of the ratchet moves with constant velocity $v = R\Omega$ along the $x$-axis. In the mirror's rest frame $x'y'z'$, the surface normal of the mirror makes an angle $\theta$ with the $z'$-axis. In the laboratory frame $xyz$, the incident plane-wave has frequency $\omega^{\mathrm{inc}}$ and propagation vector $\boldsymbol{k}^{\mathrm{inc}} = -(\omega^{\mathrm{inc}}/c)\hat{\boldsymbol{z}}$, while the reflected beam has frequency $\omega^{\mathrm{ref}}$ and wave-vector $\boldsymbol{k}^{\mathrm{ref}} = k_x^{\mathrm{ref}}\hat{\boldsymbol{x}} + k_z^{\mathrm{ref}}\hat{\boldsymbol{z}}$.

It is easy to verify that $k^{\text{ref}} = \sqrt{(k_x^{\text{ref}})^2 + (k_z^{\text{ref}})^2} = \omega^{\text{ref}}/c$. These results are valid for positive as well as negative values of $\theta$, and also for positive and negative values of $v$. Upon reflection from the mirror, the change in a single photon's energy is readily computed as follows:

$$\Delta \mathcal{E} = \hbar(\omega^{\text{ref}} - \omega^{\text{inc}}) = \hbar\gamma^2 \omega^{\text{inc}}(v/c)[(v/c) + \sin(2\theta - \varphi)]. \tag{52}$$

Prior to incidence, the photon carries a momentum $p = \hbar\omega^{\text{inc}}/c$ along the negative $z$-axis. Upon reflection, there emerges a component of momentum along the $x$-axis, whose angular momentum relative to the center of the ratchet is $R\hbar k_x^{\text{ref}}$. This change in the photon's angular momentum must be equal and opposite to the change in the angular momentum $\mathcal{L} = I\Omega$ of the ratchet—$I$ being the ratchet's moment of inertia. Given that the ratchet's rotational kinetic energy is $\mathcal{E}_K = \tfrac{1}{2}I\Omega^2 = \mathcal{L}^2/(2I)$, its acquired energy as a result of the photon impact will be

$$\Delta\mathcal{E}_K \cong 2\mathcal{L}R\hbar k_x^{\text{ref}}/(2I) = R\Omega\hbar k_x^{\text{ref}} = \hbar\gamma^2 \omega^{\text{inc}}(v/c)[(v/c) + \sin(2\theta - \varphi)]. \tag{53}$$

Clearly, the change in the kinetic energy of the ratchet is equal and opposite to the change in the photon energy upon reflection as revealed by the Doppler shift of its frequency. The blackbody radiation at temperature $T$ is thus seen to be capable of performing mechanical work on the ratchet, albeit at the expense of reducing the temperature of the radiation. The temperature drop is in the ratio of $\omega^{\text{ref}}/\omega^{\text{inc}}$.

**12. A proof of quantization of thermal radiation**. An important feature of thermal radiation interacting with material objects is intimately tied to the fluctuation-dissipation theorem [20,21]. With reference to Fig.7, imagine a highly reflective aluminum foil dropping under the pull of gravitational attraction inside an evacuated chamber containing only thermal radiation. According to the fluctuation-dissipation theorem, a friction force must be acting on the foil to oppose its acceleration under the pull of gravitation. The origin of this friction force is the Doppler shift of the various modes of the EM field inside the chamber, as seen in the foil's rest frame.

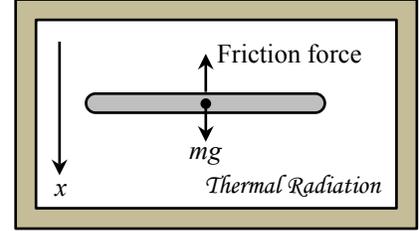

**Fig.7**. Inside an evacuated chamber filled with thermal radiation at the equilibrium temperature $T$, a highly reflective aluminum foil of mass $m$ and surface area $A$ drops under the gravitational acceleration $g$. The fluctuation-dissipation theorem requires that the foil eventually reach a steady-state velocity where the gravitational force $mg$ is cancelled by a friction force. The origin of the friction force is the Doppler shift of the various radiation modes as seen in the foil's rest frame.

In what follows, we shall derive an expression for the friction force $\boldsymbol{F}(v,T)$, and confirm that indeed it is proportional to the velocity $v$ of the foil (when $v \ll c$) and dependent on the temperature $T$. We will then argue that the discrete nature of the EM radiation (in the form of light particles – photons – that carry energy $\hbar\omega$ and linear momentum $\hbar\omega/c$ along their propagation direction) is sufficient to guarantee that a free-floating foil would reach thermal equilibrium with its radiation-filled environment. Our analysis follows in the footsteps of Einstein in his trail-blazing papers on the quantum nature of light [15-17,22-24]. While some of Einstein's arguments in these papers depend on individual atoms or molecules having discrete energy levels $\varepsilon_1$ and $\varepsilon_2$, which result in absorption and emission of EM energy in discrete packets at the frequency $\omega = (\varepsilon_2 - \varepsilon_1)/\hbar$, his other reasonings based on thermodynamics and statistical physics leave no doubt as to the quantized nature of thermal radiation. To put it differently, Einstein's description of physical phenomena such as the photoelectric effect, photoluminescence, optical absorption, and spontaneous as well as stimulated emission of radiation require only that the exchange of energy and linear momentum with matter take place via packets that are characteristic of the discrete energy levels of the atoms and molecules involved [15-17,22]; however, his examination of a highly reflective thin foil immersed in thermal radiation reveals that the radiation itself must indeed be quantized [23,24].

To evaluate the friction force $\boldsymbol{F}$ acting on a foil that moves with constant velocity $v$ along the $x$-axis (see Fig.7), we Lorentz transform the phase-factor of a plane wave having frequency $\omega$ and $k$-vector $\boldsymbol{k}^{\text{inc}} = (\omega/c)[\cos\theta\,\hat{\boldsymbol{x}} + \sin\theta\cos\varphi\,\hat{\boldsymbol{y}} + \sin\theta\sin\varphi\,\hat{\boldsymbol{z}}]$ to the rest-frame $x'y'z'$ of the mirror, as follows:

$$\boldsymbol{k}\cdot\boldsymbol{r} - \omega t = (\omega/c)[\cos\theta\, x + \sin\theta\cos\varphi\, y + \sin\theta\sin\varphi\, z] - \omega t$$
$$= (\omega/c)\{\cos\theta\,\gamma(x' + vt') + \sin\theta\cos\varphi\, y' + \sin\theta\sin\varphi\, z'\} - \omega\gamma(t' + vx'/c^2)$$
$$= (\omega/c)\{\gamma(\cos\theta - v/c)x' + \sin\theta\cos\varphi\, y' + \sin\theta\sin\varphi\, z'\} - \gamma\omega[1 - (v/c)\cos\theta]t'. \quad (54)$$

Upon reflection from the mirror, the sign of $k'_x$ is reversed. Subsequently, a second Lorentz transformation (this time back to the $xyz$ or lab frame) yields

$$\boldsymbol{k}'\cdot\boldsymbol{r}' - \omega't' = (\omega/c)[\gamma(v/c - \cos\theta)x' + \sin\theta\cos\varphi\, y' + \sin\theta\sin\varphi\, z'] - \gamma\omega[1 - (v/c)\cos\theta]t'$$
$$= (\omega/c)[\gamma^2(v/c - \cos\theta)(x - vt) + \sin\theta\cos\varphi\, y + \sin\theta\sin\varphi\, z]$$
$$\quad - \gamma^2\omega[1 - (v/c)\cos\theta](t - vx/c^2)$$
$$= (\omega/c)\{\gamma^2[2(v/c) - (1 + v^2/c^2)\cos\theta]x + \sin\theta\cos\varphi\, y + \sin\theta\sin\varphi\, z\}$$
$$\quad - \gamma^2\omega[(1 + v^2/c^2) - 2(v/c)\cos\theta]t. \quad (55)$$

The changes in $\omega$ and $k_x$ in consequence of reflection from the mirror are thus seen to be

$$\Delta\omega = \omega^{\text{ref}} - \omega^{\text{inc}} = 2\gamma^2(v/c)[(v/c) - \cos\theta]\omega,$$
$$\Delta k_x = k_x^{\text{ref}} - k_x^{\text{inc}} = 2\gamma^2[(v/c) - \cos\theta](\omega/c). \quad (56)$$

Now, for every photon colliding with the top of the mirror at an angle $\theta$ (relative to the $x$-axis), there is one bouncing off at an angle $(\pi - \theta)$ from the bottom, each imparting a momentum $\hbar\Delta k_x$ to the foil. The distance that a photon must travel to reach the foil within a short time $\Delta t$ is $[c - (v/\cos\theta)]\Delta t$. Consequently, the energy-density must be multiplied by the volume $\pm(c\cos\theta - v)A\Delta t$ of the relevant parallelepiped, then divided by the photon energy $\hbar\omega$, to yield the number of incident photons at $\theta$ and $(\pi - \theta)$ on the surface area $A$ of the foil. The net force resulting from the impact of all such photon pairs is thus the product of $-4A\gamma^2(v/c)\cos\theta$ and the relevant energy-density (at frequency $\omega$) associated with the propagation angle $\theta$. Considering that $\langle\cos\theta\rangle = \int_0^{\pi/2}\sin\theta\cos\theta\,\mathrm{d}\theta = \tfrac{1}{2}$ and that the integrated energy-density $\mathcal{E}(\omega, T)$ over all frequencies is $4\sigma T^4/c$—the Stefan-Boltzmann constant $\sigma$ is given by Eq.(11)—the average friction force acting on the foil is readily found to be

$$F_x = -8A\gamma^2(v/c^2)\sigma T^4. \quad (57)$$

When $v \ll c$, we have $\gamma \cong 1$, in which case $F_x$ is proportional to the mirror velocity $v$, as anticipated by the fluctuation-dissipation theorem.[1]

Next, let us examine our highly reflective foil which is assumed to be in thermal equilibrium with its surrounding radiation while residing in an otherwise empty chamber at temperature $T$. The situation is similar to that depicted in Fig.7, except that the gravitational force is now absent. In accordance with the equipartition theorem of statistical physics [12,13], the average kinetic energy of the foil along the $x$-axis must be $\tfrac{1}{2}m\langle v^2\rangle = \tfrac{1}{2}k_B T$; therefore, $\langle v^2\rangle = k_B T/m$. At the same time, the friction force $\boldsymbol{F}$ dissipates the foil's kinetic energy at the constant rate of $\boldsymbol{F}\cdot\boldsymbol{v}$ per second, which, for the one-dimensional motion along $x$, is given by

$$\frac{d\mathcal{E}}{dt} = \boldsymbol{F} \cdot \boldsymbol{v} \cong -8A\langle v^2/c^2\rangle \sigma T^4 = -\frac{2\pi^2 A k_B^5 T^5}{15 mc^4 \hbar^3} \quad \text{Joule/sec.} \tag{58}$$

---

[1] As an aside, note the consequences of the existence of this friction force in the context of cosmic microwave background radiation. In particular, an observer traveling with constant velocity in intergalactic space must be able to determine his/her own velocity relative to *absolute space* by measuring the friction force against the background microwave radiation, in apparent contradiction to Einstein's principle of relativity!

This loss of energy, of course, must be overcome by the kinetic energy gained by the foil in its random collisions with the surrounding (thermal) photons. Each photon momentum must be multiplied by $2\cos\theta/m$ to yield the velocity (along $x$) that is imparted to the foil upon impact. Recall that, when adding a number of independent random variables, say, $\vartheta = \sum \vartheta_n$, we will have $\langle \vartheta^2 \rangle = \sum \langle \vartheta_n^2 \rangle$ provided that $\langle \vartheta_n \rangle = 0$. Now, the variance of the photon momentum ($\wp = \hbar\omega/c$) is the same as that of the photon energy ($\mathcal{E} = \hbar\omega$) divided by $c^2$—see Eq.(33) for the variance of the energy-density $\mathcal{E}(\omega, T)$. The variance of the foil velocity along $x$ is thus $4\cos^2\theta/(m^2 c^2)$ times the variance of the corresponding EM energy. Considering that, for impacts along the direction $\theta$ on a surface area $A$ during a time interval $\Delta t$, the relevant volume of radiation is $Ac\Delta t \cos\theta$, the variance of the foil velocity becomes $4A\Delta t \cos^3\theta/(m^2 c)$ times the variance of the corresponding EM energy-density. This result must be multiplied by ½$m$ to yield the average kinetic energy of the foil, then divided by $\Delta t$ to arrive at the time-rate of increase of the foil's kinetic energy. Noting that $\langle \cos^3\theta \rangle = \int_0^{\pi/2} \sin\theta \cos^3\theta \, d\theta = ¼$, the time-rate of increase of the kinetic energy of the foil associated with photons in the frequency interval $(\omega, \omega + d\omega)$ will be $A/(2mc)$ times $\text{Var}[\mathcal{E}(\omega, T)]$ given by Eq.(33). Integration over all frequencies now yields

$$\frac{d\mathcal{E}}{dt} = \left(\frac{A}{2mc}\right) \int_0^\infty \frac{\hbar^2 \omega^4 \exp(\hbar\omega/k_B T)}{\pi^2 c^3 [\exp(\hbar\omega/k_B T) - 1]^2} d\omega = \left(\frac{A k_B^5 T^5}{2\pi^2 mc^4 \hbar^3}\right) \int_0^\infty \frac{x^4 \exp(x)}{[\exp(x) - 1]^2} dx = \frac{2\pi^2 A k_B^5 T^5}{15 mc^4 \hbar^3}. \tag{59}$$

↑ integration by parts

The time-rates of growth and decay of the kinetic energy of the foil given by Eqs.(58) and (59) are seen to be identical, thus leaving the foil in thermal equilibrium with its environment. In this way we have confirmed that the EM radiation, in consequence of its inherently discrete nature, imparts ½$k_B T$ of kinetic energy (along $x$) to the free-floating foil, in compliance with the requirements of the equipartition theorem of statistical physics.

**13. Summary and concluding remarks**. In this paper we have derived Planck's blackbody radiation formula using two different methods (Secs.2 and 9), and related the results to Stefan-Boltzmann's law of thermal radiation (Sec.3), to the radiation pressure inside an evacuated chamber at constant temperature (Sec.4), to energy-density fluctuations (Sec.7), to the entropy of photon gas at fixed temperature (Sec.8), and to Einstein's $A$ and $B$ coefficients, which pertain to spontaneous and stimulated emission of radiation (Sec.10). We have shown that a chamber filled with a homogeneous dielectric medium of refractive index $n(\omega)$ continues to emanate radiation in accordance with the Stefan-Boltzmann law (Sec.5), even though the energy-density and radiation pressure inside the chamber are modified by the presence of $n(\omega)$.

In Sec.6 we argued against the Minkowski momentum of a photon inside a dielectric-filled chamber and showed the EM momentum-density to be that given by Abraham's formula, despite the fact that the radiation pressure acting on the walls is known to be proportional to the refractive index $n(\omega)$ of the dielectric. In Sec.11 we applied the second law of thermodynamics to surfaces of varying shapes and/or differing optical properties, and demonstrated that the pressure of thermal radiation on such surfaces is independent of the shape and/or optical properties of the surface. We also showed that thermal radiation emanating from a blackbody at a given temperature $T$ is capable of performing mechanical work without violating the second law. Finally, following in the footsteps of Einstein, we showed in Sec.12 that the principles of statistical physics lead inescapably to the conclusion that thermal radiation into vacuum is quantized—in discrete packets that have energy $\mathcal{E} = \hbar\omega$ and linear momentum $\wp = \hbar\omega/c$.

It is thus clear that thermal physics has the ability to shed light on certain problems associated with radiation pressure and photon momentum. A problem that, to our knowledge, has heretofore been ignored is the mechanism by which a transparent dielectric slab (or a spherical glass bead, or any other transparent object) acquires is average kinetic energy of ½$k_B T$ per degree of freedom in accordance with the equipartition theorem. Perhaps one could invoke the Balazs thought experiment [25] in conjunction with Einstein's methods as applied to individual atoms or free-floating mirrors [17,24] to address this problem.

# Appendix A

To derive the Stefan-Boltzmann law from thermodynamic principles, let the energy $U$ residing in a chamber of volume $V$ be in thermal equilibrium with the walls. Denote the absolute temperature of the box by $T$, and let the (isotropic) pressure on the interior walls be $p$. A well-known identity, derived directly from the second law of thermodynamics (see [11], Chapter 45, Sec.1, Eq.(45.7)), is the following:

$$(\partial U/\partial V)_T = T(\partial p/\partial T)_V - p. \tag{A1}$$

Now, inside an otherwise empty box, the radiation will be in thermal equilibrium with the walls, the energy $U$ of the EM field being proportional to the volume $V$. Consequently, $(\partial U/\partial V)_T = u(T)$, where $u(T)$ is the EM energy-density of blackbody radiation at temperature $T$. Also, according to Eq.(12), $p(T) = \tfrac{1}{3}u(T)$. Therefore, Eq.(A1) may be written

$$u = (T/3)(\partial u/\partial T)_V - (u/3) \quad \rightarrow \quad 4u(T) = T(\partial u/\partial T)_V$$
$$\rightarrow \quad du/u = 4\, dT/T \quad \rightarrow \quad \ln u(T) = \ln T^4 \quad \rightarrow \quad u(T) = (4\sigma/c)T^4. \tag{A2}$$

In the final expression in Eq.(A2), the integration constant, written as $4\sigma/c$, incorporates the Stefan-Boltzmann constant $\sigma$ defined in Eq.(11). The total energy density of blackbody radiation obtained from the laws of thermodynamics thus exhibits the same $T^4$ dependence as the Planck law of Eq.(9), albeit *without* an explicit derivation of the coefficient $4\sigma/c$.

The identity in Eq.(A1) may also be derived by starting from the fundamental thermodynamic relation $\Delta S = (1/T)\Delta U + (p/T)\Delta V$, where $S(U,V,N)$ is the entropy of a closed system in thermal equilibrium with a reservoir at absolute temperature $T$, the closed system having internal energy $U$, volume $V$, and a fixed number of particles, $N$ [12]. Since the number $N$ of particles in the system is constant, we ignore the variable $N$ and attempt to express $S$, $p$ and $T$ as functions of the remaining variables $U$ and $V$. From the fundamental relation we have $1/T = (\partial S/\partial U)_V$ and $p/T = (\partial S/\partial V)_U$. Expanding $T(U,V)$ to first order in the variables $U$ and $V$, we find

$$\Delta T = \left(\frac{\partial T}{\partial U}\right)_V \Delta U + \left(\frac{\partial T}{\partial V}\right)_U \Delta V = 0 \quad \rightarrow \quad \left(\frac{\partial U}{\partial V}\right)_T = -\frac{(\partial T/\partial V)_U}{(\partial T/\partial U)_V}. \tag{A3}$$

Substituting the above identity into Eq.(A1) now yields

$$(\partial U/\partial V)_T = T(\partial p/\partial T)_V - p \quad \rightarrow \quad -(\partial T/\partial V)_U = T(\partial p/\partial T)_V (\partial T/\partial U)_V - p(\partial T/\partial U)_V$$
$$\rightarrow \quad -(\partial T/\partial V)_U = T(\partial p/\partial U)_V - p(\partial T/\partial U)_V = T^2 \partial(p/T)/\partial U|_V$$
$$\rightarrow \quad \partial(1/T)/\partial V|_U = \partial(p/T)/\partial U|_V \quad \rightarrow \quad \frac{\partial^2 S}{\partial U \partial V} = \frac{\partial^2 S}{\partial V \partial U}. \tag{A4}$$

The last equality confirms the validity of the thermodynamic identity in Eq.(A1).

**Digression**: Using the thermodynamic relations $1/T = (\partial S/\partial U)_V$ and $p/T = (\partial S/\partial V)_U$, one can prove the identity $p = T(\partial S/\partial V)_T - (\partial U/\partial V)_T$ that relates the pressure $p$ to the rates of change of the internal energy $U$ and the entropy $S$ with volume $V$ at the constant temperature $T$. This is done by expanding to first order the functions $U(V,T)$ and $S(V,T)$ in terms of the small variations in the independent variables $V$ and $T$, we find

$$\Delta U = (\partial U/\partial V)_T \Delta V + (\partial U/\partial T)_V \Delta T = 0 \quad \rightarrow \quad (\partial T/\partial V)_U = -(\partial U/\partial V)_T/(\partial U/\partial T)_V. \tag{A5}$$

$$\Delta S = (\partial S/\partial V)_T \Delta V + (\partial S/\partial T)_V \Delta T \quad \rightarrow \quad (\partial S/\partial V)_U = (\partial S/\partial V)_T + (\partial S/\partial T)_V (\partial T/\partial V)_U$$

$$\rightarrow \quad p/T = (\partial S/\partial V)_T - (\partial S/\partial T)_V (\partial U/\partial V)_T/(\partial U/\partial T)_V = (\partial S/\partial V)_T - (\partial S/\partial U)_V (\partial U/\partial V)_T$$

$$\rightarrow \quad p/T = (\partial S/\partial V)_T - (1/T)(\partial U/\partial V)_T \quad \rightarrow \quad p = T(\partial S/\partial V)_T - (\partial U/\partial V)_T. \tag{A6}$$

In terms of the Helmholtz free energy of the system, $F = U - TS$, the preceding identity is written as $p = -(\partial F/\partial V)_T$. The pressure $p$ observed under an *isothermal* expansion of the volume $V$ is thus associated with the change in the free energy $F$, as opposed to the change in the internal energy $U$. This is a consequence of the fact that a certain amount of heat, $\Delta Q$, is needed to maintain the temperature $T$ during a reversible expansion by $\Delta V$. The change of the entropy, $\Delta S = \Delta Q/T$, allows one to express the total change in the internal energy of the system as $\Delta U = T\Delta S - p\Delta V$, which leads straightforwardly to the thermodynamic identity under consideration.

# Appendix B

Here we show, for a homogeneous plane-wave in free space, that $\varepsilon_0 \boldsymbol{E}_0 \cdot \boldsymbol{E}_0 = \mu_0 \boldsymbol{H}_0 \cdot \boldsymbol{H}_0$. Assuming a plane-wave of frequency $\omega_0$ and propagation vector $\boldsymbol{k}$, Maxwell's 3$^{\text{rd}}$ equation, $\boldsymbol{\nabla} \times \boldsymbol{E} = -\partial \boldsymbol{B}/\partial t$, yields $\boldsymbol{k} \times \boldsymbol{E}_0 = \mu_0 \omega_0 \boldsymbol{H}_0$. Squaring both side of this equation, we find $(\boldsymbol{k} \times \boldsymbol{E}_0) \cdot (\boldsymbol{k} \times \boldsymbol{E}_0) = (\mu_0 \omega_0)^2 \boldsymbol{H}_0 \cdot \boldsymbol{H}_0$, which may be simplified as

$$k^2 \boldsymbol{E}_0 \cdot \boldsymbol{E}_0 - (\boldsymbol{k} \cdot \boldsymbol{E}_0)^2 = (\mu_0 \omega_0)^2 \boldsymbol{H}_0 \cdot \boldsymbol{H}_0. \tag{B1}$$

From Maxwell's 1$^{\text{st}}$ equation, $\boldsymbol{\nabla} \cdot \boldsymbol{E} = 0$, we have $\boldsymbol{k} \cdot \boldsymbol{E}_0 = 0$. Using the dispersion relation, $k^2 = (\omega_0/c)^2$, Eq.(B1) yields $(\omega_0/c)^2 \boldsymbol{E}_0 \cdot \boldsymbol{E}_0 = (\mu_0 \omega_0)^2 \boldsymbol{H}_0 \cdot \boldsymbol{H}_0$. Substituting $\mu_0 \varepsilon_0$ for $1/c^2$, we arrive at $\varepsilon_0 \boldsymbol{E}_0 \cdot \boldsymbol{E}_0 = \mu_0 \boldsymbol{H}_0 \cdot \boldsymbol{H}_0$.